%% file: Draft.tex
\documentclass[journal,comsoc, onecolumn, 12pt]{IEEEtran}
\pdfoutput=1
\usepackage{mathtools,amssymb,amsmath,amsthm} 
\usepackage[hyphens]{url}
\usepackage{hyperref} 
\usepackage{bbold}
\usepackage{graphicx} 
\usepackage{siunitx} 
\usepackage{multirow}
\usepackage[font = footnotesize]{subcaption} 
\usepackage[labelsep=period,font = footnotesize]{caption}
\usepackage{algorithm,algorithmic} 
\usepackage{cite}
\graphicspath{ {Figures/} } 
\usepackage{color}
\newtheorem{theory}{Theorem}
\newtheorem{prop}{Proposition}
\newcommand{\norm}[1]{\left\lVert#1\right\rVert_{1}}  

\begin{document}
\title{Technical Report \\ Analytical Modeling and Improvement of Interference-Coupled RAN Slicing}
\author{Seyed~Ali~Hashemian and
        Farid~Ashtiani
\thanks{The authors are with the Department of Electrical Engineering, Sharif University of Technology (SUT). Tehran 11155-4363, Iran (email: a.hashemian17@sharif.edu, ashtianimt@sharif.edu) }%
\thanks{A condensed version of this technical report has been submitted as a journal paper.}%
}
\maketitle

\begin{abstract}
The emerging 5G technology needs to support simultaneously running incompatible service types on a common infrastructure. Network slicing is a solution that corresponds a slice of the network to each service type. Ensuring that user activity in one slice does not affect other slices, i.e., inter-slice isolation, is a key requirement of slicing. Since due to interference and channel conditions, wireless link quality is unpredictable and variable, providing isolation in radio access network (RAN) is cumbersome. In this technical report, we consider multi-cell RAN slicing where the coupled interference between cells results in dynamic behavior for slices. We propose a queueing theoretic-based model to analyze interference-coupled multi-cell RAN slicing. To this end, we map our scenario on a suitable state-dependent queueing network and propose an iterative algorithm to obtain approximately the network steady-state probability distribution and derive average delay and throughput. To quantify isolation in slices, we define some new key performance indicators (KPIs). Finally, we propose and analyze an interference-aware channel allocation policy that avoids use of overlapped frequency channels for as much as possible. Numerical results demonstrate the accuracy of our proposed model and the efficacy of the interference-aware policy in improving isolation-based KPIs compared to random allocation policy.
\end{abstract}


\IEEEpeerreviewmaketitle
\clearpage
\input{intro}
\clearpage
\input{"system_model"}
\clearpage
\input{"queueing_model"}
\clearpage
\input{"kpi"}
\clearpage
\input{"allocation"}
\clearpage
\input{"numerical"}
\clearpage
\input{"conclusion"}
\input{"appendix"}

\ifCLASSOPTIONcaptionsoff
  \newpage
\fi
\newpage
\bibliographystyle{IEEEtran}
\bibliography{IEEEabrv,references}

\end{document}

%% file: intro.tex
\section{Introduction} \label{sec:intro}
\IEEEPARstart{T}{he} fifth generation of mobile networks (5G), which is now widely available in major urban areas, is by design addressing a wide range of needs of the vertical industries by offering various service types, including enhanced mobile broadband (eMBB), ultra-reliable low latency communication (URLLC), and machine-type communications (mMTC) where each one requires  its quality of service (QoS) highly different from the others \cite{5GPPP2021}. It is estimated that 660 million subscribers worldwide by the end of 2021 have employed 5G, and this number is predicted to increase to around 8.9 billion by the end of 2027 \cite{ericson2021}. The conventional one-size-fits-all network structure will not be enough to cope with the anticipated massive amounts of traffic demand and diverse types of services with often conflicting requirements, and therefore, the 5G architecture needs to be more flexible and adaptive.

 Network slicing is a solution in which the network is partitioned into separate slices where each one is configured to serve a specific type of service or users, and all these slices are simultaneously running on a common underlying infrastructure. To realize network slicing, the existing physical resources in a network need to be abstracted as virtual resources and then be assigned to slices. This process is known as wireless network virtualization (WNV). In a WNV business model, multiple logical roles can be defined as some entities owning the physical resources and letting other entities utilize them as virtual resources. In a three-layer business model, infrastructure provider (InP) owns the infrastructure equipment and spectrum resources and provides the infrastructure-as-a-service (IaaS) for tenants. Mobile virtual network operators (MVNOs) lease the network resources, create virtual resources, and then assign the virtual resources to service providers (SPs). Finally, SPs provide services for their users \cite{liang2014wireless}. 

Network slicing can be implemented at different levels, including core network (CN), transport network (TN), and RAN. In the RAN domain, spectrum resources are scarce and need to be managed in an elastic, customizable, and efficient way to satisfy all users' requirements \cite{afolabi2018network}. Similar to 4G systems, frequency resources in 5G RAN will be segmented as physical resource blocks (PRBs), where each PRB corresponds to the minimum allocation unit. However, 5G systems are more flexible as there can be multiple numerologies, i.e., multiple types of spacing for subcarriers \cite{3GPP2021}.

Isolation is a required property of network slicing that ensures performance of each tenant doesn't alter while multiple tenants use their network slices. Isolation can be done at different levels of network slicing, and depending on the slice definition, inter-slice and intra-slice isolations should be considered \cite{afolabi2018network}. The wireless nature of RAN makes effective and isolated slicing a difficult task as the frequency resources are limited and the wireless link capacity is variable. In fact, for a wireless link, the capacity is a function of signal-to-interference-plus-noise ratio (SINR) and allocated bandwidth. Therefore, changes in interference level or location of receivers affect SINR and, consequently, the link capacity \cite{richart2016resource}. Slice isolation in the RAN domain is more challenging when slices serve different types of services with heterogeneous and even conflicting QoS requirements, and user traffic is more dynamic and unpredictable. The modeling and analysis of slicing, as well as its improvement, are the main focus of this technical report. 

\subsection{Related Works}
RAN slicing is a broad area of research, and many problems can be defined there. Depending on the problem, isolation can be a concern at different levels, including physical resource scheduling (to control the SINR value) and admission control (to preserve the QoS of existing users) \cite{su2019resource}. The designing problem of slicing configuration includes whether an InP accepts a request for a slice formation and how should MVNOs rent resources and form the slices. Then there is a traffic assignment problem which includes which slice should be assigned to user traffic from a particular SP and how much traffic should be forwarded to that slice. Finally, the intra-slice resource allocation problem includes how MVNOs allocate the slice resources to the users.

Some works studied general resource allocation in network slicing as an extension to resource allocation in conventional networks, assuming that inter-slice interference has been handled in other layers. Therefore, the slice isolation concept appears in the form of satisfying the slice QoS. In \cite{Azimi_2021}, without assumption of any inter-cell interference, authors considered a two-level scheduler where the first level periodically allocates PRBs to slices and the second level allocates power and PRBs to users to satisfy their QoS and then optimized these allocations using machine learning techniques. Authors in \cite{Gonzalez_2020} considered a single-cell scenario and offered two algorithms that allocate PRBs to slices and their users using proportional fairness while keeping slices isolated by limiting the total number of allocated PRBs. Similarly, in \cite{Marabissi_2019}, authors considered a single-cell scenario and designed a two-level PRB allocation optimization problem to satisfy QoS requirement for different types of slices where the first level allocates PRBs to slices, and then in the second level, the slice allocates PRBs to users in a fair manner. 

Other works considered the inter-slice interference isolation when allocating PRBs. Usually, the slice PRB allocation, and therefore, the isolation management, is done by the InP, and then MVNOs do the user PRB allocation. In particular, \cite{D_Oro_2021, Zambianco_2020, D_Oro_2019} studied physical resource scheduling for RAN slicing in multi-cell scenarios where they considered two levels of scheduling. In the first level of scheduling, and to satisfy the service level agreement (SLA), InP allocates PRBs to MVNO slices, maximizing the number of interfering PRBs allocated to the same MVNO. Then, in the second level of scheduling, MVNO can use conventional 5G coherent and coordinated transmission techniques to mitigate the inter-cell interference. Furthermore, authors in \cite{Zambianco_2020} considered slices with different numerologies, and therefore, the inter-slice interference also includes the inter-numerological interference. In \cite{Zarandi_2021}, the authors considered a multi-cell RAN slicing scenario with mobile edge computing (MEC) and, by deriving the user delay when downloading and embedding the isolation into an optimization problem, optimized the spectrum, power, and computational resource allocation such that slice delay requirements are satisfied. The problem with such studies is that, since the InP does the slice PRB allocation, the role of InP is prominent, and the MVNO's control over their network is limited. This contradicts the essence of network slicing, which gives virtual operators enough freedom to maximize their benefit. Moreover, in order to allocate the required resources, the InP needs to predict the users' activity beforehand. This needs extra computational resources in addition to user data from virtual networks. However, since MVNOs are in competition, and their goal is to maximize their revenue, such data is private, and they are unwilling to share it. On the other hand, MVNOs might lease the resources from multiple InPs. Besides, the available resources are usually dependent on factors like the network and business conditions, the offered price by MVNOs, and available bandwidth. In other words, the slice resource allocation process by the InP has a random nature, and like a physical network operator, it is MVNO's job to manage the leased resources to satisfy QoS and the inter-slice isolation. Stochastic approaches, including queueing theory, can provide each MVNO with some performance indicators to evaluate other MVNOs activity without the need for detailed information.

Queueing theory has been used in a few works to observe slices from a stochastic point of view and obtain values like delay and throughput for slice customers. In particular, \cite{Al_Khatib_2015} models each slice as a queue with batch arrivals and batch departures where allocated PRBs to slice are customers and then tries to obtain the steady-state probability distribution using an embedded Markov chain. Authors in \cite{Al_Khatib_2019}, went further and achieved an optimum PRB allocation scheme that maximizes the network spectrum utilization. In \cite{Feng_2020}, the authors considered a single cell uplink scenario with two slices corresponding to two types of services. By modeling each user as a discrete-time queue where user packets are customers, and channel rate is the service rate, they attempted to optimize the allocated bandwidth for each slice in addition to the power and QoS of slice users such that queues stay stable and user delay doesn't exceed a maximum value. Authors in \cite{Bin_2019,Bin_1234} studied the slice formation problem with an admission control policy and offered a greedy algorithm that helps the InP to decide whether accept requests to form a new slice or not. For each type of slice, a queue for waiting requests and a queue for accepted requests are considered so that the admission control rate affects the state of the queues. These queues then have been analyzed for the patient and impatient customers. In \cite{Ha_2020}, the authors considered the user admission control problem in a network with two linked eMBB and URLLC slices and modeled each one as a queue where frequency channels are the servers and then tried to find an admission control policy that minimizes the eMBB customer blocking probability while preserving the URLLC QoS.

Since in RAN slicing, the spectrum resources are sliced and controlled by multiple tenants on multiple cells, the inter-slice interference isolation problem can be considered as a subset of the classic inter-cell interference problem. The authors in \cite{bib:958292, Klessig2014, Oehmann2014, Fehske2014, Oehmann2013, Fehske2012,7390972} focused on the coupled nature of a multi-cell network, due to inter-cell interference, which means, interference value is a function of activity in all cells and therefore, can not be analyzed separately for a single cell. The flow traffic model, which looks at user activities from a large-timescale perspective (i.e., not in the packet level) and in a continuous-time manner, has been assumed to simplify the long-term queueing theory analysis of system performance in the time domain. In \cite{bib:958292}, the authors modeled each cell as a queue with a round robin scheduler, and therefore, all spectrum resources are either inactive or busy in serving data flows. Considering such a model for each cell under investigation simplifies the problem into a two-dimensional Markov chain as each neighboring cell either has customers and causes interference or is empty and causes no interference. Next, they offered two approximate methods, namely the state aggregation and averaged interference, to solve the Markov chain. Such modeling is not applicable to a network slicing scenario as it contradicts the fact that MVNOs only have control over their own slices and not the whole spectrum. Besides, in real systems, BSs only allocate the required amount of resources (bandwidth) to each user and keep the rest for other users to increase the fairness and multiplexing gain. 

\subsection{Contributions}
In this technical report, we consider a WNV scenario including an InP, multiple MVNOs, and multiple SPs and study the dynamic behavior of network slices under the coupled-interference environment in a multi-cell network. The goal is to investigate the effect of user activity in slices defined in neighboring cells on the performance of a particular slice that is assigned to a specific MVNO. We employ a queueing theoretic approach based on the state-dependent queues to obtain the network steady-state probability distribution which then can further be used to derive delay and throughput values for each slice and the whole network. We then discuss that, rather than randomly assigning the customers to their slices, MVNOs can cooperate and follow an interference-aware channel allocation policy to control the inter-slice interference and improve the slice isolation and QoS. In summary, our main contributions are
\begin{itemize}
	\item Formulating inter-slice isolation problem in a multi-cell RAN slicing using a coupled-interference point of view
	\item Modeling the coupled slices as a network of state-dependent queues and offering an iterative algorithm to analyze the queues and extract the network steady-state probability distribution
	\item Defining slice and network KPIs to evaluate QoS and isolation level
	\item Proposing and analyzing an interference-aware channel allocation policy that requires cooperation among MVNOs
\end{itemize}
To the best knowledge of the authors, there are no detailed studies on the effect of coupled-interference on the inter-slice isolation problem in network slicing.

The rest of the technical report is organized as follows. In Section \ref{sec:system}, we present the system model as well as our assumptions, describe the slicing configuration, and formulate the isolation problem under coupled interference among slices . In Section \ref{sec:queueing}, we model the problem using a state-dependent queueing theoretic approach and devise an iterative algorithm to obtain the network steady-state probabilities. Furthermore, in Section \ref{sec:kpi}, we present KPIs to evaluate the isolation level for slice users. In Section \ref{sec:allocation}, we propose an interference-aware channel allocation algorithm to reduce the interference and increase the isolation by cooperation among MVNOs, and then we analyze the proposed algorithm using the same state-dependent queueing network approach. In Section \ref{sec:numerical}, in order to validate our model, we carry out computer simulations and compare numerical results to benchmarks. Finally, in Section \ref{sec:conclusion}, we provide concluding remarks.

%% file: system_model.tex
\section{System Model} \label{sec:system}
\subsection{Cellular Network}
\begin{figure}[!t]
	\centering \includegraphics[width=0.6\columnwidth]{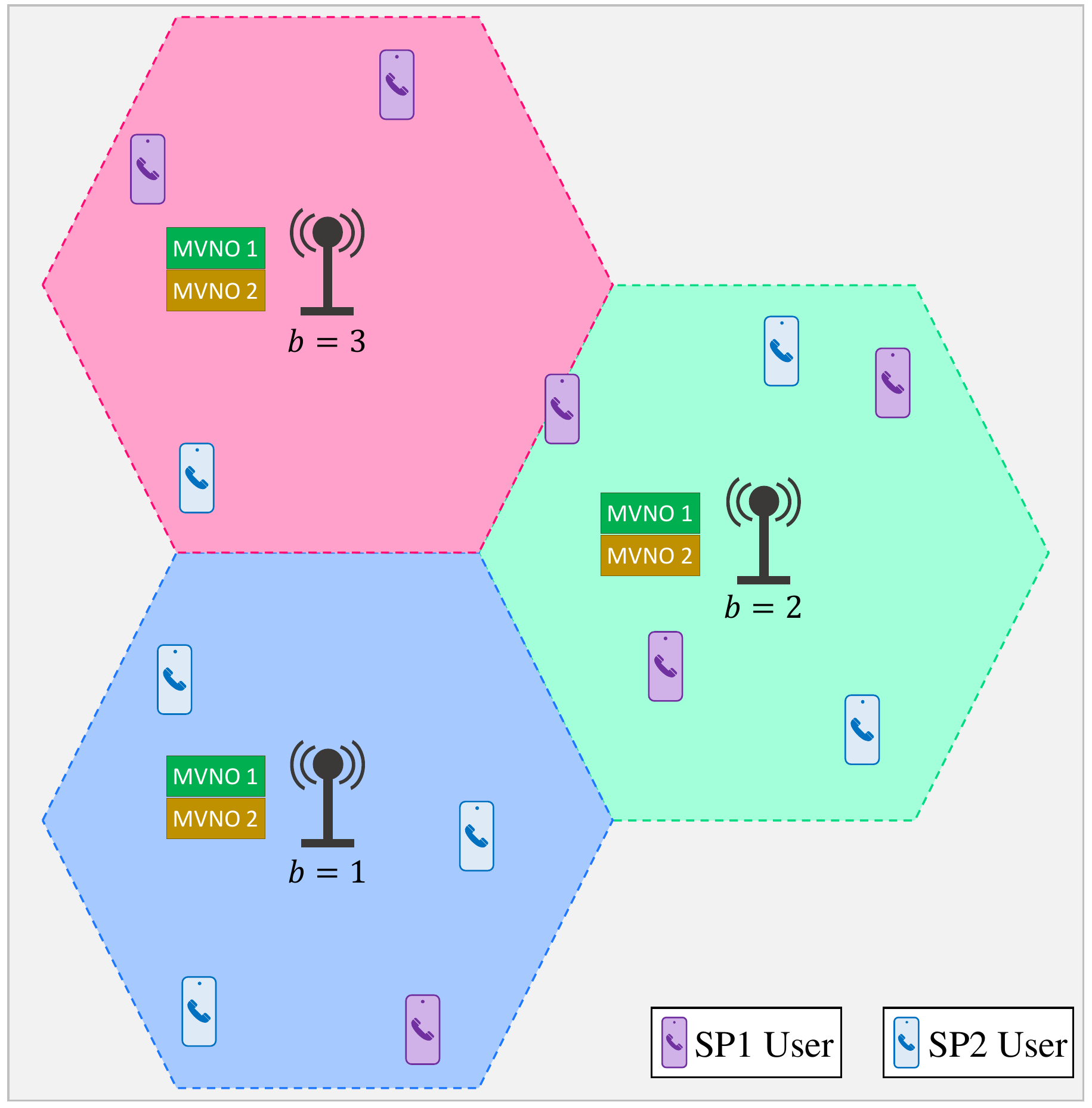}
	\caption{Illustration of the cellular network.}
	\label{fig:Cell}
\end{figure}

As shown in Fig. \ref{fig:Cell}, we consider a RAN consisting of a set $\mathcal{B} = \{1,\dots, B\}$ of base stations (BS) where BS $b \in \mathcal{B}$ has a transmission power equal to $P^{\mathrm{BS}}_{b}$ and is located at the center of its corresponding cell in a way that cell $b$ covers a region $\mathcal{L}_b \subset \mathbb{R}^2$ with no overlap. Thus, the total network coverage is $\mathcal{L} = \bigcup_{b \in \mathcal{B}} \mathcal{L}_b$. In the real world, some users can be served by multiple BSs, but here it is assumed that only the nearest BS serves each user. Here, the definition of a BS can include many RAN technologies depending on the transmission power and the cell coverage.

We consider a set $\mathcal{U} = \{1,\dots, U\}$ of SPs where SP $u \in \mathcal{U}$ provides service for a group of users with specific service requirements so that each user only receives its service from a single SP. Although it is possible to consider different simultaneous services to a user provided by some SPs, we consider such a situation as different independent users with corresponding different services. We also consider a set $\mathcal{V} = \{1,\dots, V\}$ of MVNOs where MVNO $v \in \mathcal{V}$ rents radio resources from an InP and provides service for SP users based on the corresponding SLA. We assume users are stationary or moving slowly and they don't move between different cells coverage (no handover) for the sake of simplicity. Users are distributed throughout the network coverage area and the spatial two-dimensional density of SP $u$-users at location $l \in \mathcal{L}$ is $\sigma_{u}(l)$ with $\sum_{u \in \mathcal{U}}\int_{l \in \mathcal{L}} \sigma_{u}(l)\;\mathrm{d}l = 1$. We consider a downlink scenario where SP $u$-users initiate downloading data flows at random times following a Poisson point process with rate $\lambda_{u}$ and continue until the flow has been downloaded completely. A data flow of SP $u$-users is a group of consecutive data packets that has a random number of bits with a memory-less distribution and an average $\Omega_u$. This is because, in reality, rather than transferring individual data packets within the time-scattered groups of PRBs, users usually start sessions of data downloading at random times when they want to download data files with random sizes. We also assume that a change in the flow transmission rate may happen, but regarding the memory-less assumption on the number of data bits in a flow, it is equivalent to a retransmission with the new rate. Furthermore, since the flow download time is relatively large compared to the usual channel coherence times, we may ignore the effect of fast fading as a time-average effect on flow transmission.  We do not consider the effect of shadowing on the received power for the sake of simplicity. Therefore, we assume that during the flow transmission, the average signal strength remains constant. In addition, since we consider a flow traffic model, we ignore the slotted structure of PRBs in time domain and assume each flow to be a continuous-time stream of a random number of data bits. 

\subsection{Slicing Configuration}
Let us consider that frequency spectrum at cell $b$ has been divided into a set $\mathcal{S}_b$ with $S_b = |\mathcal{S}_b|$ arbitrary, distinctive and sporadic chunks, each one called as a slice. Each slice can be considered for a specific type of service. Frequency range and the number of slices can vary in various cells and the definition of a slice is limited within the corresponding cell, i.e., for $b, b' \in \mathcal{B}$ we have $\mathcal{S}_b \cap \mathcal{S}_{b'} = \emptyset$. In other words, it is possible that two slices at different cells have some overlapped frequency bandwidths, but slices at different cells have different indices. In the entire network we have a set $\mathcal{S} = \bigcup_{b\in\mathcal{B}} \mathcal{S}_b$ of slices where each slice $s \in \mathcal{S}$ is composed of $Q_s$ frequency channels. Each channel has a bandwidth $w_s$ equivalent to a specific number of arbitrary PRBs, allocated to a single user in an OFDMA cellular network as a whole. To elaborate, when a slice channel is active, its corresponding PRBs simultaneously transmit data for the assigned user as long as a flow is being downloaded. Regarding the type of service for which a typical slice has been configured, the channels at different slices may have different frequency bandwidths. 
 
\begin{figure}[!t]
	\centering \includegraphics[width=0.75\columnwidth]{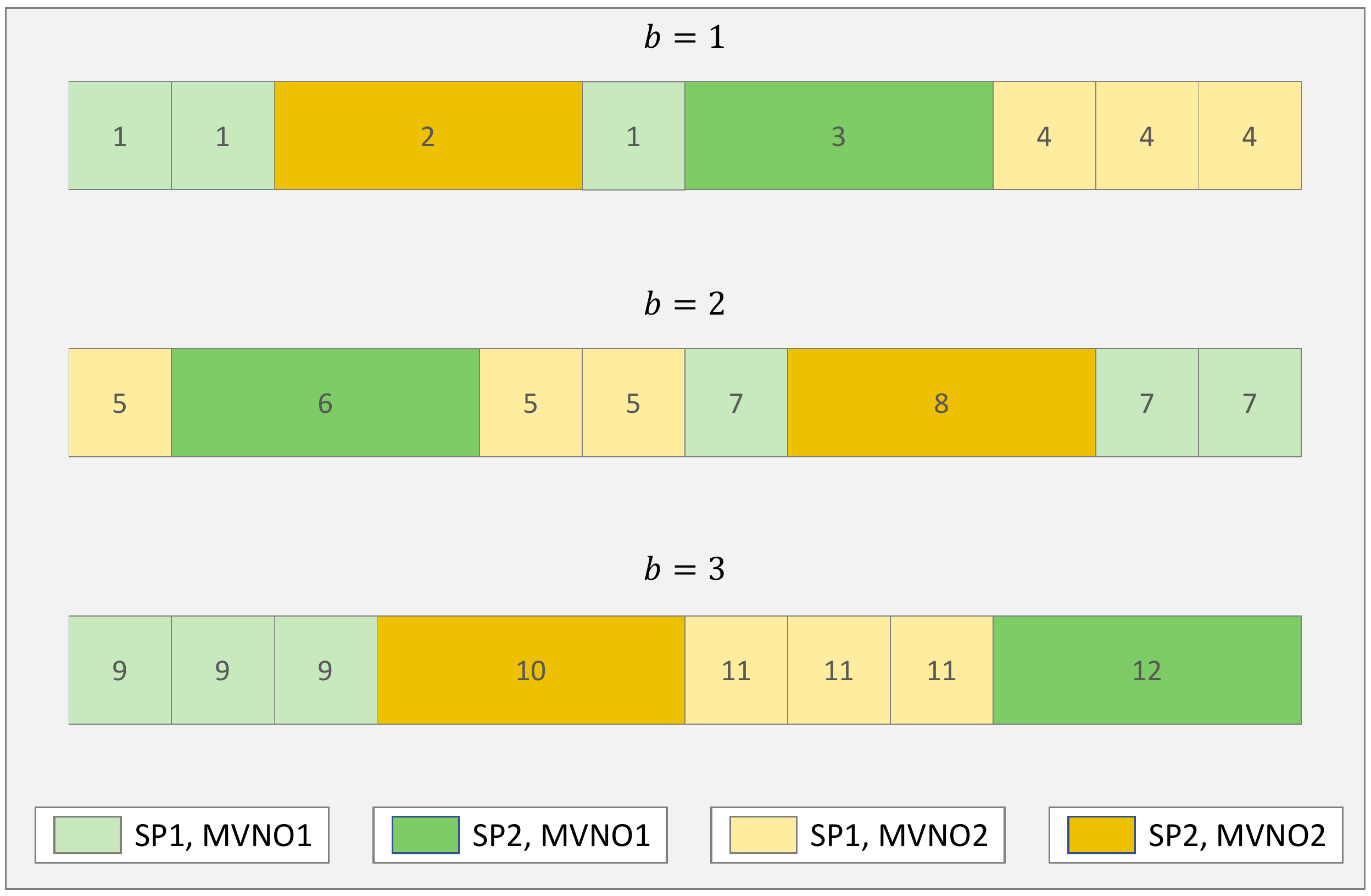}
	\caption{Sample slicing configuration. Both SP $1$ and $2$ distribute their traffic among MVNO $1$ and $2$ and MVNOs exploit their slices in different cells to support traffics.}
	\label{fig:Slice}
\end{figure}

 Each SP can distribute its users flow traffic among multiple MVNOs. Then, if possible, each MVNO assigns each flow to a channel on a proper slice. MVNOs can have different number of slices in different cells and correspond them to flows of different service types. Consequently, we define $\mathcal{S}_{b,v} \subseteq \mathcal{S}_b$ as the set of slices in cell $b$ that has been rented by MVNO $v$ with $S_{b,v} = |\mathcal{S}_{b,v}|$. We refer to each manner of slice assignment as a slicing configuration. In this technical report, we will not investigate the designing problem of slicing configuration. Instead, we will analyze the performance of a typical slicing configuration by analytical modeling of all interactions among slices. Fig. \ref{fig:Slice} depicts a sample slicing configuration with $\mathcal{S} = \{1,\dots,12 \}$, $\mathcal{S}_{1} = \{1, 2, 3, 4 \}$, $\mathcal{S}_{1,1} = \{1, 3\}$ and $\mathcal{S}_{1,2} = \{2, 4 \}$.
\subsection{Isolation Among Slices}
Isolation is a key feature of effective network slicing which means user activity on one slice should not affect other slices. However, flows assigned to slices in adjacent cells will experience interference if being transmitted on channels with some overlapped frequencies. $\mathcal{N}_{s,q}$ is defined as the set of slice-channel pairs which have frequency overlap with channel $q$ on slice $s \in \mathcal{S}$ including $(s,q)$. In other words, for all $s \in \mathcal{S}$ and $0\leq q \leq Q_{s}$ we have
\begin{equation} 
	\label{eq:316141}
	\mathcal{N}_{s,q}=\Bigl\{(s',q') \bigm\vert  s' \in \mathcal{S}, 0\leq q' \leq Q_{s'}, \substack{(s,q) \text{ and } (s',q') \\ \text{ have frequency}  \\ \text{ overlap}} \Bigr\}.	
\end{equation}

We define interference vector $\mathbf{\Delta}_{s,q} \in \{0,1\}^{|\mathcal{N}_{s,q}| - 1}$ as a binary vector so that element $\mathbf{\Delta}_{s,q}(s',q')$ is $1$ or $0$ corresponding to whether channel $(s',q') \in \mathcal{N}_{s,q} \setminus (s,q)$ is active (i.e., a data flow is allocated to $(s',q')$) or not. With this in mind, SINR for a tagged flow, i.e., a flow under investigation, which is being transmitted by BS $b$ on channel $q$ of slice $s \in \mathcal{S}_{b}$ towards location $l \in \mathcal{L}_{b}$ under interference vector $\mathbf{\Delta}_{s,q}$ can be written as
\begin{equation} 
	\label{eq:919125}
	\gamma_{s,q}(l,\mathbf{\Delta}_{s,q})=\frac {P^{\mathrm{SI}}_{s,q}(l)}{\displaystyle\sum_{(s',q') \in \mathcal{N}_{s,q} \setminus (s,q)} \mathbf{\Delta}_{s,q}(s',q') P^{\mathrm{IN}}_{(s',q'),(s,q)}(l) + N_{0}}
\end{equation}
where $P^{\mathrm{SI}}_{s, q}(l)$ is the signal power received from $b \in \mathcal{B}$ on channel $q$ of $s \in \mathcal{S}_{b}$ at $l \in \mathcal{L}_{b}$ and $P^{\mathrm{IN}}_{(s',q'),(s,q)}(l)$ is the interference power affecting $(s,q)$ received from $b' \in \mathcal{B}$ on channel $q'$ of $s' \in \mathcal{S}_{b'}$ at $l \in \mathcal{L}_{b}$ and finally $N_0$ is the additive noise power. Note that \eqref{eq:919125} is general and can include different RAN technologies as $P^{\mathrm{SI}}_{s, q}(l)$ and $P^{\mathrm{IN}}_{(s',q'),(s,q)}(l)$ are independent of RAN technology. Afterward, ignoring the limitation of user processing power, we can obtain wireless link capacity as
\begin{equation} 
	\label{eq:735027} 
	C_{s,q}(l,\mathbf{\Delta}_{s,q})=\eta_1 w_{s} log_{2}(1+\eta_2 \gamma_{s,q}(l,\mathbf{\Delta}_{s,q}))
\end{equation}
where $\eta_1$ and $\eta_2$ are bandwidth and SINR efficiencies, respectively \cite{Mogensen2007}. Since each channel consists of a group of PRBs that simultaneously transmit data, the capacity of each channel is equal to the total capacity of PRBs, and therefore \eqref{eq:735027} is justified.

By harmonic averaging \eqref{eq:735027} on all possible locations $l \in \mathcal{L}_{b}$ for SP $u$-users \cite{bib:958292}, we can achieve interference-dependent average wireless link capacity for SP $u$-user flows being transmitted on $(s , q)$ as
\begin{equation} 
	\label{eq:174806}
	C_{s,q,u}^{-1}(\mathbf{\Delta}_{s,q})=\int_{l \in \mathcal{L}_{b}} C_{s,q}^{-1}(l,\mathbf{\Delta}_{s,q})\sigma_{u}(l \mid b)\mathrm{d}l
\end{equation}
where $\sigma_{u}(l \mid b) = \frac{\sigma_{u}(l)}{\int_{l \in \mathcal{L}_{b}} \sigma_{u}(l)\mathrm{d}l}$ is the SP $u$-users density at location $l$ given that $l \in \mathcal{L}_{b}$. 

Eq. \eqref{eq:174806} implies that the average wireless link capacity for SP $u$-user flows is a function of interference from slices on adjacent cells. On the other hand, the probability of meeting interference during transmissions in a cellular network is a function of average wireless link capacity on different cells since lower link capacity means more transmission time and therefore, a higher probability of overlap in the time domain. Thereupon, average wireless link capacity is coupled among pairs of network slices and cannot be simply obtained. To guarantee the inter-slice isolation, this problem needs to be investigated in detail. That is, we need to exploit an analytical model that is able to consider the effect of dynamic coupled interference leading to variable transmission rate during downloading a flow. In fact, one can guarantee the required QoS for each slice by knowing the average wireless link capacity under coupled interference and allocating enough resources. Furthermore, by understanding the interference effect on the wireless link capacity and having performance metrics to evaluate isolation, one can control the inevitable inter-slice interference and increase the isolation using a suitable spectrum allocation. For this purpose, we use queueing theoretic analysis to model and study coupled interactions among slices. 

%% file: queueing_model.tex
\section{Queueing Theoretic Analysis} \label{sec:queueing}
In this section, we propose a queueing theoretic model to address the interference coupling problem and analyze network performance metrics. 

\subsection{Flow Traffic and Queue Model}
\begin{figure*}[!h]
	\centering \includegraphics[width=\textwidth]{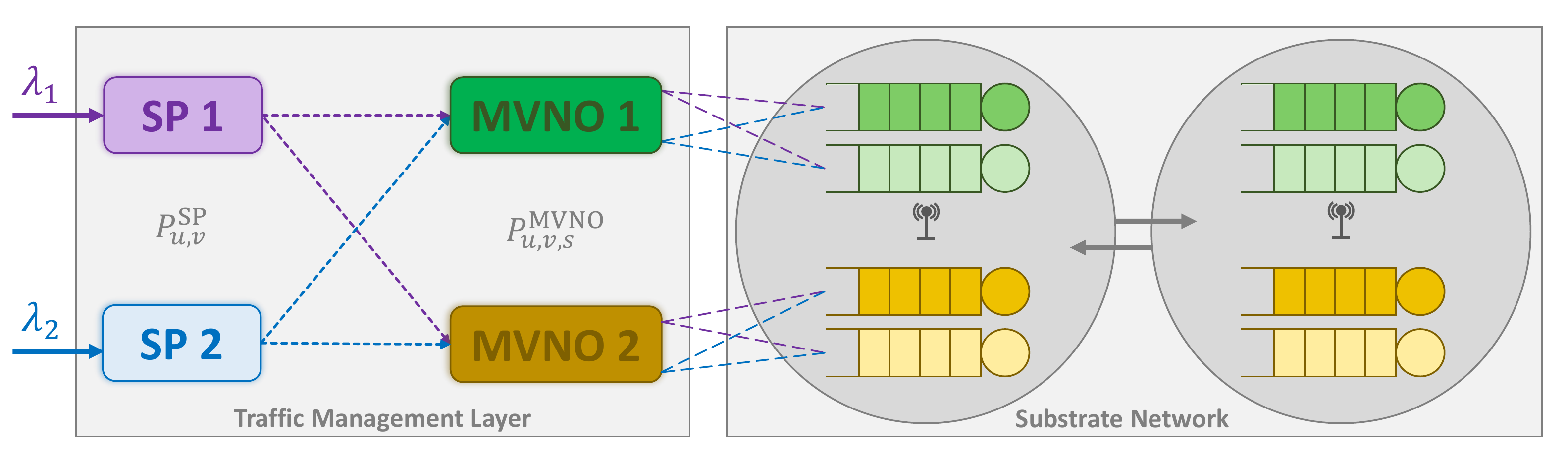}
	\caption{SPs and MVNOs can manage user generated flow traffic using aforementioned parameters. The traffic then arrives at queues embedded in substrate cells with coupled wireless link capacity.}
	\label{fig:Queue}
\end{figure*}

As we mentioned in Section \ref{sec:system}, users of SPs generate independent and identically distributed (i.i.d.) data flow traffic at random times following a Poisson point process such that the flow rate for SP $u$-users is $\lambda_{u}$. Then, based on some contracts between SPs and MVNOs, traffic of the SP users needs to be handled by the corresponding MVNOs. As shown in Fig. \ref{fig:Queue}, to control the network performance and ensure QoS for different types of services, SPs and MVNOs are able to manage the distribution of their data flow traffic. For each SP $u$, we define $P_{u,v}^{\mathrm{SP}}$ with $\sum_{v \in \mathcal{V}} P_{u,v}^{\mathrm{SP}} = 1$ as the probability that SP $u$ delivers its flow traffic to MVNO $v$. Moreover, for $s \in \mathcal{S}_{b,v}$, we define $P_{u,v,s}^{\mathrm{MVNO}}$ with $\sum_{s \in \mathcal{S}_{b,v}} P_{u,v,s}^{\mathrm{MVNO}} = 1$ as the probability that MVNO $v$ assigns a flow of SP $u$ to slice $s$. The values of $P_{u,v}^{\mathrm{SP}}$ and $P_{u,v,s}^{\mathrm{MVNO}}$ depend on SPs and MVNOs policies. Afterward, for slice $s \in \mathcal{S}_{b, v}$, we can achieve the arrival rate of users' flows from SP $u$ as
\begin{equation} 
	\label{eq:174443}
	\lambda_{s, u} = \lambda_{u} P_{u,v}^{\mathrm{SP}} \sigma_{u, b} P_{u,v,s}^{\mathrm{MVNO}}
\end{equation}
where $\sigma_{u, b} = \int_{l \in \mathcal{L}_{b}} \sigma_{u}(l) \mathrm{d}l$ is the probability that SP $u$-users are in cell $b$. 

We consider each slice $s \in \mathcal{S}$ as a queue where data flows are customers and channels are servers. Then hereafter, we will use the terms \textit{slice} and \textit{queue} interchangeably. Customers from SP $u$ arrive at queue $s$ with rate $\lambda_{s, u}$ as a Poisson process and stay in queue until they receive service, i.e., being completely transmitted on one of $Q_s$ channels of the corresponding slice. The service time for each flow depends on its number of bits and wireless link capacity. Therefore, regarding interference coupling among slices and other assumptions, we consider a negative exponential distribution with variable rates. This means that we have a network of queues with service rate for each queue depending on the state of the other queues (i.e., the existing status of each channel) in the network. If the number of arrived customers is more than the existing servers, extra customers are buffered until we have a total of $Q^{\mathrm{max}}_s$ customers in the queue. If the buffer capacity is full, newly arrived customers would be blocked. Accordingly, in the entire system, we would have a network of queues where each one has a standard $M/M(n)/Q_s/Q^{\mathrm{max}}_s$ model. Since the state of each queue depends on the state of the whole network, we propose a state-dependent queueing network to model our system.
\subsection{Our Proposed State-Dependent Queueing Network Model}
We say $s, s' \in \mathcal{S}$ interact with each other, when they either have an overlapped frequency range or there exists at least one slice $s'' \in \mathcal{S}$ that has some overlapped distinct frequency ranges with both $s$ and $s'$. For example, in Fig. \ref{fig:Slice}, slices $s = 7$ and $s = 8$ interact with each other because they both have overlapped frequency range with $s = 12$. Keeping this in mind, one can partition $\mathcal{S}$ into disjoint subsets where queues in each subset interact with each other but not with queues in other subsets. Then, queues in each subset can be analyzed independently. Without loss of generality, we assume that all queues in $\mathcal{S}$ interact with each other.

Now consider a two-part and non-negative vector $\mathbf{n} = (\mathbf{n}_{\mathrm{LH}}, \mathbf{n}_{\mathrm{RH}})$ of the state space $\mathbb{N}_{\mathbf{n}}$ as the network state vector where $\mathbf{n}_{\mathrm{LH}}$ and $\mathbf{n}_{\mathrm{RH}}$ are the left-hand and the right-hand parts with equal size $|\mathcal{S}|$. For each queue $s \in \mathcal{S}$, the elements $\mathbf{n}_{\mathrm{RH}}(s)$ and $\mathbf{n}_{\mathrm{LH}}(s)$, respectively, represent the number of customers and the remaining buffer capacity and therefore we have $\mathbf{n}_{\mathrm{LH}}(s) + \mathbf{n}_{\mathrm{RH}}(s) = Q^{\mathrm{max}}_{s}$. Also consider a similar two-part and non-negative vector $\mathbf{a} = (\mathbf{a}_{\mathrm{LH}}, \mathbf{a}_{\mathrm{RH}})$ of the state space $\mathbb{N}_{\mathbf{a}}$ as the movement vector where for each queue $s \in \mathcal{S}$, the elements $\mathbf{a}_{\mathrm{RH}}(s)$ and $\mathbf{a}_{\mathrm{LH}}(s)$, respectively, represent the number of departures from and arrivals to $s$. Regarding the continuous time dynamic in our scenario only single movements are possible at each time. Therefore, for all vectors $\mathbf{a}$, the first norm is equal to one, i.e., $\lVert \mathbf{a} \rVert_{1}  = 1$. As a numerical example, for slicing configuration shown in Fig. \ref{fig:Slice} with $|\mathcal{S}| = 12$, following vectors correspond to arrivals at queue $s = 1$ when there are no customers in the network and each queue has a unity-size buffer (i.e., $Q^{\mathrm{max}}_{s} = Q_{s} + 1$)
\begin{equation} 
	\begin{split}
		\mathbf{n} &= (\overbrace{4, 2, 2, 4, 4, 2, 4, 2, 4, 2, 4, 2}^{\mathbf{n}_{\mathrm{LH}}}, \overbrace{0, 0, 0, 0, 0, 0, 0, 0, 0, 0, 0, 0}^{\mathbf{n}_{\mathrm{RH}}})\\
		\mathbf{a} &= (\overbrace{1, 0, 0, 0, 0, 0, 0, 0, 0, 0, 0, 0}^{\mathbf{a}_{\mathrm{LH}}}, \overbrace{0, 0, 0, 0, 0, 0, 0, 0, 0, 0, 0, 0}^{\mathbf{a}_{\mathrm{RH}}}).
	\end{split}
	\nonumber
\end{equation}

For each possible pair of $(\mathbf{n}, \mathbf{a})$ vectors, we define $\xi(\mathbf{n}, \mathbf{a})$ as the rate of leaving the state $\mathbf{n}$ through the movement $\mathbf{a}$. Vector $\mathbf{a}$, then, transforms into another vector $\mathbf{a}'$ with probability $r(\mathbf{a}, \mathbf{a}')$ which changes the network state from $\mathbf{n}$ to $\mathbf{n}'$. If $\mathbf{a}$ corresponds to an arrival to $s$, $\mathbf{a}'$ corresponds to a departure from $s$, and vice versa. We define $\langle \mathbf{n}, \mathbf{a} \rangle$ as the set of all pairs of $(\mathbf{n}, \mathbf{a})$ that can transform into each other, i.e., $\mathbf{n} - \mathbf{a} = \mathbf{n}' - \mathbf{a}'$. In the previous example, vector $\mathbf{a}$ transforms into $\mathbf{a}' = (0, 0, 0, 0, 0, 0, 0, 0, 0, 0, 0, 0, 1, 0, 0, 0, 0, 0, 0, 0, 0, 0, 0, 0)$ with $r(\mathbf{a}, \mathbf{a}') =1$ changing the network state from $\mathbf{n}$ to $\mathbf{n}' = (3, 2, 2, 4, 4, 2, 4, 2, 4, 2, 4, 2, 1, 0, 0, 0, 0, 0, 0, 0, 0, 0, 0, 0)$. Obviously, in our model at each $\langle \mathbf{n}, \mathbf{a} \rangle$ just two members exist. We also define $\pi(\mathbf{n})$ as the network steady-state probability distribution, i.e., the normalized network stationary measure. Now we use the following theorem from \cite{book:Miyazawa}.

\begin{theory}\label{theory:951737}
	The network stationary measure $\tilde{\pi}$ satisfies the local balance equations
	\begin{equation}
		\tilde{\pi}(\mathbf{n})\xi(\mathbf{n},\mathbf{a}) = \sum_{(\mathbf{n}',\mathbf{a}') \in \langle \mathbf{n}, \mathbf{a} \rangle}\tilde{\pi}(\mathbf{n}')\xi(\mathbf{n}',\mathbf{a}')r(\mathbf{a}', \mathbf{a})
	\end{equation}
	if and only if there exist a non-negative function $\Psi$ and a positive function $\Phi$ such that for all possible pairs of $(\mathbf{n}, \mathbf{a})$ the rate $\xi(\mathbf{n},\mathbf{a})$ take the form
	\begin{equation}
	\xi(\mathbf{n},\mathbf{a}) = \frac{\Psi(\mathbf{n} - \mathbf{a})}{\Phi(\mathbf{n})}.
	\end{equation}	
	When this is true, we have
		\begin{equation}
		\tilde{\pi}(\mathbf{n}) = \Phi(\mathbf{n})
	\end{equation}
\end{theory}
\begin{proof}
	Please see chapter 9 in \cite{book:Miyazawa}.
\end{proof}

In general, definition of $P_{u,v,s}^{\mathrm{MVNO}}$ implies that each slice can serve customers from multiple SPs. But in this technical report, we assume $P_{u,v,s}^{\mathrm{MVNO}}$ value is such that slice $s\in \mathcal{S}$ at each time has only customers from a single SP $u$. This assumption is only for the sake of simplicity in formulations and can be relaxed by defining multiple new states within each slice corresponding to each SP. We would try to keep the formulation general, but we may occasionally use the notation $u_s$ to remind that $u$ depends on $s$. If we name the customers from each SP a \textit{class} of customers, then the latter assumption means that we consider queues with a single class of customers. Note that the terms \textit{type of service} and \textit{class of customer} refer to two separate concepts, and there can be SPs that offer the same type of service. Therefore, for queue $s$ which has only customers from SP $u$, we define $\Lambda_{s,u}$ and $\mathrm{M}_{s,u}$ as the state-dependent arrival and service rates considering that $\mathrm{M}_{s,u}$, due to the coupled interference, is dependent on the network state, i.e., right-hand part $\mathbf{n}_{\mathrm{RH}}$, while $\Lambda_{s,u}$ is only dependent on the state of $s$, i.e., $\mathbf{n}_{\mathrm{RH}}(s)$.
Now if we define functions $\Phi$ and $\Psi$ in a way that for each $s \in \mathcal{S}$ we have
\begin{equation} 
	\label{eq:778126}
	\xi(\mathbf{n}, \mathbf{a}) = \frac{\Psi (\mathbf{n} - \mathbf{a})}{\Phi (\mathbf{n})} = 
	\begin{cases}
		\Lambda_{s,u}(\mathbf{n}_{\mathrm{RH}}(s))\;, & \;\; \mathbf{a}_{\mathrm{LH}}(s) =1 \\
		\mathrm{M}_{s,u}(\mathbf{n}_{\mathrm{RH}})\;, & \;\; \mathbf{a}_{\mathrm{RH}}(s) =1 \\
	\end{cases},
\end{equation}
then as a corollary of Theorem \ref{theory:951737}, $\Phi (\mathbf{n})$ is a stationary measure of the network. For this purpose, we present the following proposition.

\begin{prop}\label{prop:111737}
Consider a two-part and non-negative variable vector $\mathbf{x} = (\mathbf{x}_{\mathrm{LH}}, \mathbf{x}_{\mathrm{RH}})$ similar to $\mathbf{n}$. The functions
\begin{equation} 
	\label{eq:237045}
	\Phi (\mathbf{x}) = \overbrace{\prod_{s = 1}^{|\mathcal{S}|} \prod_{x = \mathbf{x}_{\mathrm{RH}}(s)}^{Q^{\mathrm{max}}_s - 1} \Lambda_{s,u_{s}}^{-1}(x)}^{\Phi_1(\mathbf{x})} \overbrace{\prod_{\substack{s = 1, \\ \mathbf{x}_{\mathrm{RH}}(s) \neq 0, \\ e_s \neq 1}}^{|\mathcal{S}|} \mathrm{M}_{s,u_{s}}^{-1}(\mathbf{x}_{\mathrm{RH}})}^{\Phi_2(\mathbf{x})}
\end{equation}
and
\begin{equation} 
	\label{eq:300065}
	\Psi (\mathbf{x}) = \overbrace{\Big[\Phi_1(\mathbf{x}+\mathbf{e}(\mathbf{x})) \prod_{s=1}^{|\mathcal{S}|}\Big(1 - (1 - \Lambda_{s,u_{s}}(Q^{\mathrm{max}}_s))\delta(1 + \mathbf{x}_{\mathrm{LH}}(s))\Big) \Big]}^{\Psi_1(\mathbf{x})} \times \overbrace{\Big[\Phi_2(\mathbf{x})\Big]_{\mathbf{x}\rightarrow \mathbf{x} + \mathbf{e}(\mathbf{x})}}^{\Psi_2(\mathbf{x})}
\end{equation}
satisfy \eqref{eq:778126} when the rate $\xi(\mathbf{n}, \mathbf{a})$ correspond to a departure, i.e., $\mathbf{a}_{\mathrm{RH}}(s) =1$. 
\end{prop}
\begin{proof}
	Please see Appendix \ref{app1}.
\end{proof}

In proposition \ref{prop:111737}, $\delta(.)$ is the Dirac delta function and $\mathbf{e}(\mathbf{x})$ is a two-part vector-valued function indicating whether there has been a movement in each queue or not and we have 
\begin{equation} 
	\label{eq:175203}
	\mathbf{e}(\mathbf{x}) = (\mathbf{0}_{\mathrm{LH}}, \mathbf{e}_{\mathrm{RH}}(\mathbf{x}))
\end{equation}
where the left-hand part $\mathbf{0}_{\mathrm{LH}}$ is a zero vector with size $|\mathcal{S}|$ and the right-hand part $\mathbf{e}_{\mathrm{RH}}(\mathbf{x}) = (e_s)_{s \in \mathcal{S}}$ is a vector with size $|\mathcal{S}|$ and for element $e_s$ we have
\begin{equation} 
	\label{eq:979248}
	e_s = 
	\begin{cases}
		1 & \; \; \mathbf{x}_{\mathrm{RH}}(s) + \mathbf{x}_{\mathrm{LH}}(s) \neq Q^{\mathrm{max}}_s \\
		0 & \; \; \text{o.w.}
	\end{cases}.
\end{equation}
 Finally, notation $[\Phi_2(\mathbf{x})]_{\mathbf{x}\rightarrow \mathbf{x} + \mathbf{e}(\mathbf{x})}$ implies that one must determine $\Phi_2(\mathbf{x})$ first and then substitute all values of $\mathbf{x}$ with $\mathbf{x} + \mathbf{e}(\mathbf{x})$. The general idea here is to use the function $\mathbf{e}(\mathbf{x})$ to model perfectly all departure movements (but not the arrival movements) in \eqref{eq:300065} since a distinction between arrival and departure movements exists in \eqref{eq:778126}. We explain this matter in the sequel.
 
\subsection{State-Dependent Service Rate}
To determine $\mathrm{M}_{s,u}(\mathbf{n}_{\mathrm{RH}})$, we need to consider that for each realization of $\mathbf{n}_{\mathrm{RH}}$, multiple channel allocations are possible resulting in multiple values for $\mathbf{\Delta}_{s,q}$ (interference vector corresponding to $(s,q)$). By averaging $C^{-1}_{s,q,u}(\mathbf{\Delta}_{s,q})$ (see \eqref{eq:174806}) over all possible values of $\mathbf{\Delta}_{s,q}$ we have
\begin{equation} 
	\label{eq:286539}
	C_{s,q,u}^{-1}(\mathbf{n}_{\mathrm{RH}})=
	\sum_{\mathbf{\Delta}_{s,q}\in \{0,1\}^{|\mathcal{N}_{s,q}| - 1}} \mathrm{Pr}(\mathbf{\Delta}_{s,q})C_{s,q,u}^{-1}(\mathbf{\Delta}_{s,q})
\end{equation}
where $\mathrm{Pr}(\mathbf{\Delta}_{s,q})$ is the probability that the interference vector $\mathbf{\Delta}_{s,q}$ occurs. MVNOs can assign flows to vacant channels with different policies leading to different $\mathrm{Pr}(\mathbf{\Delta}_{s,q})$. In the case of random and independent channel allocation policy in all queues, we have
	\begin{equation} 
		\label{eq:990352}
		\mathrm{Pr}(\mathbf{\Delta}_{s,q})= \prod_{s' \in \hat{\mathcal{N}}_{s,q} \setminus (s)} \mathrm{Pr}(\mathbf{\Delta}_{s,q,s'}),
	\end{equation}
	where $\hat{\mathcal{N}}_{s,q} = \{s' | \exists q', (s',q') \in \mathcal{N}_{s,q}\}$ is the set of slices which have frequency overlap with $(s, q)$ and $\mathbf{\Delta}_{s,q,s'}$ is a sub-vector of $\mathbf{\Delta}_{s,q}$ corresponding to the interference from channels of slice $s'$, and we have
	\begin{equation} 
		\label{eq:255651}
			\mathrm{Pr}(\mathbf{\Delta}_{s,q,s'})  = 
			\begin{cases}
				1; & \substack{\dim(\mathbf{\Delta}_{s,q,s'}) = \norm{\mathbf{\Delta}_{s,q,s'}}, \\ Q^{\mathrm{max}}_{s'} \geq \mathbf{n}_{\mathrm{RH}}(s') > Q_{s'}} \vspace{0.3cm}\\
				\frac{\binom{Q_{s'} - \dim(\mathbf{\Delta}_{s,q,s'})}{\mathbf{n}_{\mathrm{RH}}(s') - \norm{\mathbf{\Delta}_{s,q,s'}}}}{\binom{Q_{s'}}{\mathbf{n}_{\mathrm{RH}}(s') }}; & \substack{\mathbf{n}_{\mathrm{RH}}(s') - \norm{\mathbf{\Delta}_{s,q,s'}} \geq 0, \\ \mathbf{n}_{\mathrm{RH}}(s') - \norm{\mathbf{\Delta}_{s,q,s'}} \leq Q_{s'} - \dim(\mathbf{\Delta}_{s,q,s'}) } \vspace{0.3cm} \\
				0; &   \text{o.w.}
			\end{cases}
	\end{equation}
	where $\dim(.)$ and $\norm{.}$, respectively, represent the dimension and the first norm of a vector. In justifying \eqref{eq:255651}, it is worth noting that it is possible that more than one channel of a typical slice $s'$ have frequency overlap with $(s,q)$.

Since the slices are arbitrary and in general, frequency range overlaps between two slices are asymmetric, channels of a single slice may have different average wireless link capacities indicated by $C_{s,q,u}(\mathbf{n}_{\mathrm{RH}})$ in \eqref{eq:286539}. In terms of queueing theory, this means that we have queues with heterogeneous servers with varying service rates, which are complex to analyze. Therefore, for simplicity, we study an equivalent homogeneous queue with a total service rate equal to that of the heterogeneous queue. In other words, the service rate of each server in equivalent queue is the average of the heterogeneous queue service rates. For this purpose, in each state $\mathbf{n}_{\mathrm{RH}}$, considering a tagged flow that is assigned to arbitrary channel $q$ of slice $s$ and averaging $C_{s,q,u}(\mathbf{n}_{\mathrm{RH}})$ over all channels of slice $s$, we have
\begin{equation} 
	\label{eq:669958}
	C_{s,u}(\mathbf{n}_{\mathrm{RH}})=
	\sum_{q = 1}^{Q_s}\mathrm{Pr}(q)C_{s,q,u}(\mathbf{n}_{\mathrm{RH}})
\end{equation}
as the wireless link capacity for each channel in the equivalent homogeneous slice (queue), where $\mathrm{Pr}(q)$ is the probability that channel $q$ is assigned to a transmitting tagged flow. In the case of uniformly random channel allocation scheme, we have $\mathrm{Pr}(q) = 1/Q_s$. Afterwards, we can achieve the state-dependent service rate for SP $u$-customers by dividing the obtained wireless link capacity by the average number of bits of the flows as in the following:
\begin{equation} 
	\label{eq:516771}
	\mathrm{M}_{s, u}(\mathbf{n}_{\mathrm{RH}})=
	\begin{cases}
		\frac{C_{s, u}(\mathbf{n}_{\mathrm{RH}})}{\Omega_{u}}\min(\mathbf{n}_{\mathrm{RH}}(s), Q_s) ,  &0<\mathbf{n}_{\mathrm{RH}}(s)\leq Q^{\mathrm{max}}_s\\
		1, &\text{o.w.}
	\end{cases}
\end{equation}
where the term $\frac{C_{s, u}(\mathbf{n}_{\mathrm{RH}})}{\Omega_{u}}$ represents the rate of serving flows at each server of queue $s$ and the term $\min(\mathbf{n}_{\mathrm{RH}}(s), Q_s)$ implies that the service rate for queue $s$ increases as the number of customers $\mathbf{n}_{\mathrm{RH}}(s)$ increases until we have $Q_s$ customers or more, in which case, all servers are busy. When there are zero customers in a queue, the service rate is meaningless. Therefore, in order to neutralize the effect of $\mathrm{M}_{s, u}(\mathbf{n}_{\mathrm{RH}})$ in \eqref{eq:237045} and \eqref{eq:300065}, we define $\mathrm{M}_{s, u}(\mathbf{n}_{\mathrm{RH}}) = 1$ in all such states.

So far and for a known network state vector $\mathbf{n}$, service rate $\mathrm{M}_{s, u}(\mathbf{n}_{\mathrm{RH}})$ has been obtained by multiple averaging over $C_{s,q,u}(\mathbf{\Delta}_{s,q})$, regarding the fact that in each network state, channel allocation for each slice is random and independent of the state of other slices. However, when investigating the steady-state probability distribution for a specific queue, the coupling effect among slices is a burden. In fact, to achieve the steady-state probability distribution, one must know the service rate for all states of queue, but, as indicated in $\mathrm{M}_{s, u}(\mathbf{n}_{\mathrm{RH}})$, service rate is a function of the state of other network queues. Thus, we need to solve the equations recursively.

\subsection{State-Dependent Arrival Rate}
We define state-dependent arrival rate as
\begin{equation} 
	\label{eq:073213}
	\Lambda_{s,u}(\mathbf{n}_{\mathrm{RH}}(s))=
	\begin{cases}
		\lambda_{s, u} & \; \; 0\leq \mathbf{n}_{\mathrm{RH}}(s) < Q^{\mathrm{max}}_{s}  \\
		0 &\; \; \text{o.w.}
	\end{cases}.
\end{equation}
Obviously, due to the consideration of blocking, when there are $Q^{\mathrm{max}}_{s}$ customers in a queue, the arrival rate to the queue is zero. 

The defined $\Phi$ and $\Psi$ do not satisfy \eqref{eq:073213} perfectly. In fact, by substituting state-dependent rates defined by $\eqref{eq:516771}$ and $\eqref{eq:073213}$ in $\eqref{eq:237045}$ and $\eqref{eq:300065}$, one can investigate that $\eqref{eq:778126}$ will have the following form
\begin{equation}
	\label{eq:897009}
	\xi(\mathbf{n}, \mathbf{a}) = \frac{\Psi (\mathbf{n} - \mathbf{a})}{\Phi (\mathbf{n})} = 
	\begin{cases}
		\Lambda_{s,u}(\mathbf{n}_{\mathrm{RH}}(s))\beta_{\mathbf{n},\mathbf{a}} & \;\; \mathbf{a}_{\mathrm{LH}}(s) =1 \\
		\mathrm{M}_{s,u}(\mathbf{n}_{\mathrm{RH}}) & \;\; \mathbf{a}_{\mathrm{RH}}(s) =1 \\
	\end{cases},
\end{equation}
where $\beta_{\mathbf{n},\mathbf{a}}$ is a coefficient showing up due to $\Phi$ and $\Psi$ imperfection when the pair $(\mathbf{n}, \mathbf{a})$ corresponds to a permitted arrival. 

For arrivals to slice $s \in \mathcal{S}$ with $0 \leq \mathbf{n}_{\mathrm{RH}}(s)<Q^{\mathrm{max}}_s$ and from \eqref{eq:237045} and \eqref{eq:300065} we obtain
\begin{equation} 
	\label{eq:588041}
	\beta_{\mathbf{n},\mathbf{a}} =\; \mathrm{M}_{s, u}(\mathbf{n}_{\mathrm{RH}}) \prod_{s' \neq s, n_{s'}>0} \frac{\mathrm{M}_{s', u_{s'}}(\mathbf{n}_{\mathrm{RH}})}{\mathrm{M}_{s', u_{s'}}(\mathbf{n}_{\mathrm{RH}} + \mathbf{1}_{s})} 
\end{equation}
where $\mathbf{1}_{s}$ is a vector with size $|\mathcal{S}|$ that its $s\text{-th}$ element is equal to one and the rest are equal to zero.

To neutralize the effect of $\beta_{\mathbf{n},\mathbf{a}}$, a solution is to introduce a modified state-dependent arrival rate $\hat{\Lambda}_{s,u}(\mathbf{n}_{\mathrm{RH}}(s))$ such that $\hat{\Lambda}_{s,u}(\mathbf{n}_{\mathrm{RH}}(s))\beta_{\mathbf{n},\mathbf{a}} \cong \Lambda_{s,u}(\mathbf{n}_{\mathrm{RH}}(s))$. Therefore, we define
\begin{equation} 
	\label{eq:053953}
	\hat{\Lambda}_{s,u}(\mathbf{n}_{\mathrm{RH}}(s))= 
	\begin{cases}
		\lambda_{s, u}/\hat{\beta}_{s, \mathbf{n}_{\mathrm{RH}}(s)} & \; \; 0\leq \mathbf{n}_{\mathrm{RH}}(s) < Q^{\mathrm{max}}_{s}  \\
		0 &\; \; \text{o.w.}
	\end{cases}
\end{equation}
where $\hat{\beta}_{s, \mathbf{n}_{\mathrm{RH}}(s)}$ is the modification factor. Hence, for all pairs of $(\mathbf{n},\mathbf{a})$ that correspond to an arrival to slice $s \in \mathcal{S}$ when it has $\mathbf{n}_{\mathrm{RH}}(s)$ customers, we can approximate \eqref{eq:778126} with \eqref{eq:897009} by choosing $\hat{\beta}_{s, \mathbf{n}_{\mathrm{RH}}(s)}$ so that $\lambda_{s,u}\frac{\beta_{\mathbf{n},\mathbf{a}}}{\hat{\beta}_{s, \mathbf{n}_{\mathrm{RH}}(s)}}$ has the least weighted squared error with respect to $\lambda_{s,u}$. In other words, to find the value of $\hat{\beta}_{s, \mathbf{n}_{\mathrm{RH}}(s)}$ for each $s \in \mathcal{S}$ and $0 \leq \mathbf{n}_{\mathrm{RH}}(s) \textless Q^{\mathrm{max}}_s$, we have
\begin{equation}
	\label{eq:583071}
	\min_{\hat{\beta}_{s, \mathbf{n}_{\mathrm{RH}}(s)}} \sum_{\substack{\mathbf{x} \in \mathbb{N}_{\mathbf{n}},\mathbf{a} \in \mathbb{N}_{\mathbf{a}} \\ \mathbf{a}_{\mathrm{LH}}(s) =  1\\ \mathbf{x}_{\mathrm{RH}}(s)= \mathbf{n}_{\mathrm{RH}}(s)}} \hat{\Phi}_{s, \mathbf{n}_{\mathrm{RH}}(s)}(\mathbf{x}) (1 - \frac{\beta_{\mathbf{x},\mathbf{a}}}{\hat{\beta}_{s, \mathbf{n}_{\mathrm{RH}}(s)}})^2
\end{equation}
where $\mathbf{x}$ is a variable vector similar to $\mathbf{n}$ and
\begin{equation}
	\label{eq:583371}
	\hat{\Phi}_{s, \mathbf{n}_{\mathrm{RH}}(s)}(\mathbf{x}) = \frac{\Phi(\mathbf{x})}{\sum\limits_{\substack{\mathbf{x'} \in \mathbb{N}_{\mathbf{n}} \\   \mathbf{x'}_{\mathrm{RH}}(s)= \mathbf{n}_{\mathrm{RH}}(s) }} \Phi(\mathbf{x'})}
\end{equation}
is the normalized weight for state $\mathbf{x}$ as $\Phi(\mathbf{x})$ is a stationary measure of the network according to Theorem \ref{theory:951737}.

Since $\Phi(\mathbf{x})$ is dependent on $\hat{\beta}_{s, \mathbf{n}_{\mathrm{RH}}(s)}$ and vice versa, an iterative approach can be used to obtain $\Phi(\mathbf{x})$ as stated in Algorithm \ref{alg:beta}. For the proof please see Appendix \ref{app2}. To make reading easier, for the rest of the technical report, we represent the condition under sigma in \eqref{eq:583071} as $\mathrm{COND}(\mathbf{x},\mathbf{a}) = \{\mathbf{x} \in \mathbb{N}_{\mathbf{n}}, \mathbf{a} \in \mathbb{N}_{\mathbf{a}}, \mathbf{a}_{\mathrm{LH}}(s) =  1, \mathbf{x}_{\mathrm{RH}}(s)= \mathbf{n}_{\mathrm{RH}}(s)\}$. 

 \begin{algorithm}[!t]
	\caption{Obtaining network steady-state probability}
	\label{alg:beta}
	\begin{algorithmic}[1]
		\STATE  Initialize $\hat{\beta}_{s, \mathbf{n}_{\mathrm{RH}}(s)}^{(1)} = 1$ for all $s \in \mathcal{S}$ and $0 \leq \mathbf{n}_{\mathrm{RH}}(s) \textless Q^{\mathrm{max}}_s$
		\STATE  Initialize \( k \leftarrow 1 \)
		\REPEAT
		\STATE Calculate for all possible $(\mathbf{n}, \mathbf{a})$
		\begin{equation}
			\nonumber
			\hat{\Lambda}^{(k)}_{s,u}(\mathbf{n}_{\mathrm{RH}}(s))= 
			\begin{cases}
				\lambda_{s, u}/\hat{\beta}_{s, \mathbf{n}_{\mathrm{RH}}(s)}^{(k)} &  \; 0\leq \mathbf{n}_{\mathrm{RH}}(s) < Q^{\mathrm{max}}_{s}  \\
				0 & \; \text{o.w.}
			\end{cases}
		\end{equation}
		\STATE Calculate for all possible $(\mathbf{n}, \mathbf{a})$
		\begin{equation}
			\nonumber
			\Phi^{(k)}(\mathbf{n}) = \prod_{s = 1}^{|\mathcal{S}|} \prod_{n = \mathbf{n}_{\mathrm{RH}}(s)}^{Q^{\mathrm{max}}_s-1} (\hat{\Lambda}_{s,u}^{(k)}(n))^{-1} \prod_{\substack{s = 1, \\ \mathbf{n}_{\mathrm{RH}}(s) \neq 0, \\ e_s \neq 1}}^{|\mathcal{S}|} \mathrm{M}_{s,u}^{-1}(\mathbf{n}_{\mathrm{RH}})
		\end{equation}
		\STATE Update for all $s \in \mathcal{S}$ and $0 \leq \mathbf{n}_{\mathrm{RH}}(s) \textless Q^{\mathrm{max}}_s$
		\begin{equation} \nonumber
			\hat{\beta}^{(k+1)}_{s, \mathbf{n}_{\mathrm{RH}}(s)} = \frac{\sum_{\mathrm{COND}(\mathbf{x},\mathbf{a})} 	\hat{\Phi}_{s, \mathbf{n}_{\mathrm{RH}}(s)}^{(k)}(\mathbf{x})\beta_{\mathbf{x},\mathbf{a}}^2}{\sum_{\mathrm{COND}(\mathbf{x},\mathbf{a})} \hat{\Phi}_{s, \mathbf{n}_{\mathrm{RH}}(s)}^{(k)}(\mathbf{x})\beta_{\mathbf{x},\mathbf{a}}}
		\end{equation}
		\STATE \( k \leftarrow k + 1 \)
		\UNTIL{convergence}
		\RETURN the latest $\Phi^{(k)}(\mathbf{n})$ for all possible network states $\mathbf{n}$
	\end{algorithmic} 
\end{algorithm}

For each network state $\mathbf{n}$, the obtained $\Phi(\mathbf{n})$ using Algorithm \ref{alg:beta} is an approximate stationary measure of the network. The accuracy of this approximation depends on the weighted squared error in \eqref{eq:583071} which is related to three factors: the number of terms in summation, the weights $\hat{\Phi}_{s, \mathbf{n}_{\mathrm{RH}}(s)}(\mathbf{x})$ and the values of $\beta_{\mathbf{x},\mathbf{a}}$. 

An increase in the number of possible network states means more possible pairs of $(\mathbf{x},\mathbf{a})$ in \eqref{eq:583071}, leading to larger error value. The value of $\hat{\Phi}_{s, \mathbf{n}_{\mathrm{RH}}(s)}(\mathbf{x})$ can vary depending on the network traffic as it is a normalized stationary measure of the network. In light traffic, the weights for states with a low number of customers are dominant and other weights are almost negligible. In contrast, in heavy traffic, the weights for states with a large number of customers are dominant and other weights are almost negligible. In other words, for light and heavy traffic, the number of terms in summation \eqref{eq:583071} is fewer than the one in moderate traffic and therefore, the error value is less. Our numerical results confirm this matter.

To investigate the third factor, if we call $\hat{\beta}^{*}_{s, \mathbf{n}_{\mathrm{RH}}(s)}$ as the value that has been obtained from the last iteration of Algorithm \ref{alg:beta}, we can calculate the minimum cost. For arrivals to slice $s \in \mathcal{S}$ with $0 \leq \mathbf{n}_{\mathrm{RH}}(s) \textless Q^{\mathrm{max}}_s$, by substituting $\hat{\beta}^{*}_{s, \mathbf{n}_{\mathrm{RH}}(s)}$ into \eqref{eq:583071} we have

\begin{equation}
\label{eq:214038}
\begin{split} 
	\sum_{\mathrm{COND}(\mathbf{x},\mathbf{a})} \hat{\Phi}_{s, \mathbf{n}_{\mathrm{RH}}(s)}(\mathbf{x}) (1 - \frac{\beta_{\mathbf{x},\mathbf{a}}}{\hat{\beta}^{*}_{s, \mathbf{n}_{\mathrm{RH}}(s)}})^2 &= \sum_{\mathrm{COND}(\mathbf{x},\mathbf{a})} \hat{\Phi}_{s, \mathbf{n}_{\mathrm{RH}}(s)}(\mathbf{x}) (1 - \frac{\beta_{\mathbf{x},\mathbf{a}}\sum_{\mathrm{COND}(\mathbf{x'},\mathbf{a'})} \hat{\Phi}_{s, \mathbf{n}_{\mathrm{RH}}(s)}(\mathbf{x'})\beta_{\mathbf{x'},\mathbf{a'}}}{\sum_{\mathrm{COND}(\mathbf{x'},\mathbf{a'})} \hat{\Phi}_{s, \mathbf{n}_{\mathrm{RH}}(s)}(\mathbf{x'})\beta_{\mathbf{x'},\mathbf{a'}}^2})^2	 \nonumber \\
	&= 1 - \frac{\big(\sum_{\mathrm{COND}(\mathbf{x},\mathbf{a})} \hat{\Phi}_{s, \mathbf{n}_{\mathrm{RH}}(s)}(\mathbf{x})\beta_{\mathbf{x},\mathbf{a}}\big)^2}{\sum_{\mathrm{COND}(\mathbf{x},\mathbf{a})} \hat{\Phi}_{s, \mathbf{n}_{\mathrm{RH}}(s)}(\mathbf{x})\beta_{\mathbf{x},\mathbf{a}}^2} = 1- \frac{\mathrm{E}^2[\beta_{\mathbf{x},\mathbf{a}}]}{\mathrm{E}[\beta^{2}_{\mathbf{x},\mathbf{a}}]} =\frac{\mathrm{Var}[\beta_{\mathbf{x},\mathbf{a}}]}{\mathrm{E}[\beta^{2}_{\mathbf{x},\mathbf{a}}]}	  \nonumber
\end{split}
\end{equation}
where the symbols  $\mathrm{E}[\beta_{\mathbf{x},\mathbf{a}}] = \sum_{\mathrm{COND}(\mathbf{x},\mathbf{a})} \hat{\Phi}_{s, \mathbf{n}_{\mathrm{RH}}(s)}(\mathbf{x})\beta_{\mathbf{x},\mathbf{a}}$ and $\mathrm{Var}[\beta_{\mathbf{x},\mathbf{a}}] = \mathrm{E}[\beta^{2}_{\mathbf{x},\mathbf{a}}] - \mathrm{E}^{2}[\beta_{\mathbf{x},\mathbf{a}}]$, respectively, represent the expectation and the variance. Consequently, any factor that increases the ratio between $\mathrm{Var}[\beta_{\mathbf{x},\mathbf{a}}]$ and $\mathrm{E}[\beta^{2}_{\mathbf{x},\mathbf{a}}]$, increases the weighted squared error in \eqref{eq:583071}.

%% file: kpi.tex
\section{Network KPIs} \label{sec:kpi}
In this section, using the network steady-state probability distribution $\pi(\mathbf{n})$ obtained in Section \ref{sec:queueing}, we extract the network KPIs.
\subsection{Slice Steady-State Probability Distribution}
We define $\pi_{s}$ as the steady-state probability distribution for slice $s$ where $\pi_{s}(n)$ indicates the steady-state probability distribution of $n$ customers being present at queue $s$ and we have
\begin{equation} 
	\label{eq:833565}
	\pi_{s}(n) = \sum_{\substack{\mathbf{n} , \\  \mathbf{n}_{\mathrm{RH}}(s) = n}} \pi(\mathbf{n}).
\end{equation}
We also define the blocking probability $P^{B}_{s}$ as the probability of customers arriving at queue $s$ being blocked and we have
\begin{equation} 
	\label{eq:609347}
	P^{B}_{s} = \pi_{s}(Q^{\mathrm{max}}_{s}).
\end{equation}
\subsection{Slice Throughput and Delay}
We define $T_{s, u}$ as the throughput for SP $u$-users in slice $s$ and we have
\begin{equation} 
	\label{eq:557550}
	T_{s, u}=\lambda_{s,u}(1 - P^{B}_{s})\Omega_{u}.
\end{equation}
We also define $D_{s, u}$ as the average total delay for SP $u$-users in slice $s$. Using the Little's law we have
\begin{equation} 
	\label{eq:149347}
	D_{s, u}=\frac{\sum_{n = 0}^{Q_{s}^{\mathrm{max}}} n\pi_{s}(n)}{\lambda_{s,u}(1 - P^{B}_{s})}
\end{equation}
where the numerator indicates the average number of SP $u$-users in queue $s$ and the denominator indicates the rate of SP $u$-users admitted in queue $s$.

For each customer entering queue $s$, its total delay consists of sojourn time in buffer and service time. We define $D^{\mathrm{SJ}}_{s, u}$ and $D^{\mathrm{SV}}_{s, u}$, respectively, as the average sojourn and service time for SP $u$-users in queue $s$. With an approach similar to \eqref{eq:149347} we have 
\begin{equation} 
	\label{eq:709147}
	D^{\mathrm{SJ}}_{s, u} =  	\frac{\sum_{n = Q_{s}}^{Q_{s}^{\mathrm{max}}} (n -  Q_{s})\pi_{s}(n)}{\lambda_{s,u}(1 - P^{B}_{s})}
\end{equation}

\begin{equation} 
	\label{eq:123456}
	D^{\mathrm{SV}}_{s, u} = \frac{\sum_{n = 0}^{Q_{s}^{\mathrm{max}}} \mathrm{min}(n, Q_{s})\pi_{s}(n)}{\lambda_{s,u}(1 - P^{B}_{s})}
\end{equation}
where the numerators, respectively, indicate the average number of SP $u$-users in buffer and servers of queue $s$, and the denominators, respectively, indicate the rate of SP $u$-users entering buffer and servers of queue $s$. One can verify that, for SP $u$-users in queue $s$, the equation $D_{s, u} = D^{\mathrm{SJ}}_{s, u} + D^{\mathrm{SV}}_{s, u}$ holds as expected.
\subsection{Network Throughput and Delay}
Regardless of slice and cell, each MVNO has to serve SP $u$-users based on an SLA in the entire network. We define $T^{\mathrm{MVNO}}_{u, v}$ as the network throughput for SP $u$-users being served by MVNO $v$ and we have
\begin{equation} 
	\label{eq:946610}
	T^{\mathrm{MVNO}}_{u, v} = \sum_{b\in\mathcal{B}} \sum_{s \in \mathcal{S}_{b,v}} T_{s, u}.
\end{equation}

Smillar to the slice delay, we define $D^{\mathrm{MVNO}}_{u, v}$, $D^{\mathrm{MVNO, SJ}}_{u, v}$ and $D^{\mathrm{MVNO, SV}}_{u, v}$, respectively, as the network total delay, network sojourn time and network service time for SP $u$-users being served by MVNO $v$ and we have
\begin{equation} 
	\label{eq:429037}
	D^{\mathrm{MVNO}}_{u, v} = \sum_{b\in \mathcal{B} }\sum_{s \in \mathcal{S}_{b,v}}\sigma_{u, b} P_{u,v,s}^{\mathrm{MVNO}} D_{s, u}
\end{equation}
\begin{equation} 
	\label{eq:138348}
	D^{\mathrm{MVNO, SJ}}_{u, v} = \sum_{b\in \mathcal{B} }\sum_{s \in \mathcal{S}_{b,v}}\sigma_{u, b} P_{u,v,s}^{\mathrm{MVNO}} D^{\mathrm{SJ}}_{s, u}
\end{equation}
\begin{equation} 
	\label{eq:802243}
	D^{\mathrm{MVNO, SV}}_{u, v} = \sum_{b\in \mathcal{B} }\sum_{s \in \mathcal{S}_{b,v}}\sigma_{u, b} P_{u,v,s}^{\mathrm{MVNO}} D^{\mathrm{SV}}_{s, u}
\end{equation}
and the equation $D^{\mathrm{MVNO}}_{u, v} = D^{\mathrm{MVNO, SJ}}_{u, v} + D^{\mathrm{MVNO, SV}}_{u, v}$ holds.
\subsection{Isolation Metrics}
We already stated the importance of isolation among slices. To quantify and compare isolation level in the network, here we propose some metrics. Suppose that the isolation for slices that are allocated to SP $u$-users being served by MVNO $v$ is under investigation and another SP $u'$-users traffic has been continuously increased from $\lambda_{u'}$ to $\lambda_{u'} + \Delta\lambda$. Since the network KPIs are functions of SP $u'$-users traffic, a KPI versus interfering traffic curve can be obtained. Therefore, for network total delay, we can define average of delay deviation (ADD) and variance of delay deviation (VDD) as
\begin{equation} 
	\label{eq:341110}
	\mathrm{ADD}(D^{\mathrm{MVNO}}_{u, v}, u') = \frac{1}{\Delta\lambda} \int_{z= \lambda_{u'}}^{\lambda_{u'} + \Delta\lambda} \frac{D^{\mathrm{MVNO}}_{u, v}}{D^{\mathrm{MVNO},\mathrm{ZI}}_{u, v}} \mathrm{d}z
\end{equation}

\begin{equation} 
	\label{eq:941150}
	\mathrm{VDD}(D^{\mathrm{MVNO}}_{u, v}, u')  =  \frac{1}{\Delta\lambda} \int_{z= \lambda_{u'}}^{\lambda_{u'} + \Delta\lambda} \bigg(\frac{D^{\mathrm{MVNO}}_{u, v}}{D^{\mathrm{MVNO},\mathrm{ZI}}_{u, v}} - \mathrm{ADD}(D^{\mathrm{MVNO}}_{u, v}, u')\bigg)^2 \mathrm{d}z
\end{equation}
where $D^{\mathrm{MVNO},\mathrm{ZI}}_{u, v}$ is the network total delay for SP $u$-users being served by MVNO $v$ when there is zero interference in the network.
In \eqref{eq:341110}, first, the value of $D^{\mathrm{MVNO}}_{u, v}$ is normalized to zero-interference delay, so it indicates the deviation from the ideal condition as a result of the interference from SP $u'$-users traffic. Then, the average of the normalized delay versus interfering traffic curve (i.e., the area under the curve divided by the traffic changes) is stated as ADD. Furthermore, for VDD, the variance of the normalized curve would be used. Similarly, for network throughput we can define average of throughput deviation (ATD) and variance of throughput deviation (VTD) as
\begin{equation} 
	\mathrm{ATD}(T^{\mathrm{MVNO}}_{u, v}, u') = \frac{1}{\Delta\lambda} \int_{z= \lambda_{u'}}^{\lambda_{u'} + \Delta\lambda} \frac{T^{\mathrm{MVNO},\mathrm{ZI}}_{u, v}}{T^{\mathrm{MVNO}}_{u, v}} \mathrm{d}z
\end{equation}

\begin{equation} 
	\mathrm{VTD}(T^{\mathrm{MVNO}}_{u, v}, u') = \frac{1}{\Delta\lambda} \int_{z= \lambda_{u'}}^{\lambda_{u'} + \Delta\lambda} \bigg(\frac{T^{\mathrm{MVNO},\mathrm{ZI}}_{u, v}}{T^{\mathrm{MVNO}}_{u, v}} - \mathrm{ATD}(T^{\mathrm{MVNO}}_{u, v}, u')\bigg)^2 \mathrm{d}z
\end{equation}
where $T^{\mathrm{MVNO},\mathrm{ZI}}_{u, v}$ is the network throughput for SP $u$-users being served by MVNO $v$ when there is zero interference in the network.  Finally, we can combine the effects of users traffic from all other SPs and define the following metrics
\begin{align} 
\label{eq:349410}
\mathrm{ADD}(D^{\mathrm{MVNO}}_{u, v}) &= \frac{1}{U-1}\sum_{u' \neq u} \mathrm{ADD}(D^{\mathrm{MVNO}}_{u, v}, u')\\
	\mathrm{VDD}(D^{\mathrm{MVNO}}_{u, v}) &= \frac{1}{U-1}\sum_{u' \neq u} \mathrm{VDD}(D^{\mathrm{MVNO}}_{u, v}, u')\\
	\mathrm{ATD}(T^{\mathrm{MVNO}}_{u, v}) &= \frac{1}{U-1}\sum_{u' \neq u} \mathrm{ATD}(T^{\mathrm{MVNO}}_{u, v}, u') \\
	\mathrm{VTD}(T^{\mathrm{MVNO}}_{u, v}) &= \frac{1}{U-1}\sum_{u' \neq u} \mathrm{VTD}(T^{\mathrm{MVNO}}_{u, v}, u').
\end{align}

ADD, ATD determine, on average, how far the KPI is from the ideal condition, i.e., zero-interference, and VDD, VTD determine how much the KPI varies with respect to the average values. Both average and variance metrics are dimensionless and a lower value means more isolation as they both quantify the effect of changes in interfering traffic on the network KPIs. Ideally and in a fully isolated network, average and variance are equal to 1 and 0, respectively.

%% file: allocation.tex
\section{Proposing an Interference-aware Slice Channel Allocation} \label{sec:allocation}
Equations \eqref{eq:990352} and \eqref{eq:255651} were obtained assuming a uniformly random and independent channel allocation in each slice. Provided that each BS can measure the interference from other BSs, state-dependent framework presented in Section \ref{sec:queueing} can be used to analyze more complex channel allocation policies. Here we propose an interference-aware channel allocation policy aiming to decrease interference level and keep the slices more isolated.

First, an allocation lookup table $LT$ is constructed such that interference among slices is at a minimum level. For each pair $(s, q)$ in network state $\mathbf{n}$, binary value $LT(\mathbf{n}, s, q) \in \{0,1\}$ indicates whether channel $q$ of $s$ can be allocated to flows or not. Next, lookup table $LT$ is distributed among MVNOs and is valid as long as the slicing configuration is not changed. Assuming cooperation among MVNOs to mitigate the interference and know the network state $\mathbf{n}$, an MVNO that owns slice $s$ is only allowed to allocate flows to channel $q$ if $LT(\mathbf{n}, s, q) = 1$. After identifying the permitted channels, MVNO randomly assigns flows to those channels.

To construct the lookup table and for each state $\mathbf{n}$, we prioritize slices $s \in \mathcal{S}$ according to the value of $\mathbf{n}_{\mathrm{RH}}(s)/Q_s$, i.e., slices with a higher ratio of customers to total channels have higher priority. The motivation behind such prioritization is less flexibility in channel allocation in such slices. In the case of some slices having equal ratio of customers to total channels, we randomly prioritize them over each other.

Having the priority for each slice, we start by an all zero $LT$ and obtain values of $LT(\mathbf{n}, s, q)$ for all slices $s \in \mathcal{S}$ in the order of their priorities. For this purpose, we calculate the average wireless link capacity for all channels belonging to each slice one at a time to choose the best ones. In other words, for all channels $1\leq q \leq Q_{s}$ in slice $s$, we calculate $C_{s,q,u}(\mathbf{\Delta}_{s,q})$ using \eqref{eq:174806} ignoring the interference from slices with lower priority and assuming that slice $s$ is only under the interference from slices with higher priority, i.e., all interfering channels that so far have value 1 in $LT$. Therefore, interference vector $\mathbf{\Delta}_{s,q}$ in \eqref{eq:174806} can be obtained from the last values in $LT$. 

Next, we rank slice $s$ channels from highest to lowest according to the calculated $C_{s,q,u}(\mathbf{\Delta}_{s,q})$ and we choose the first $\mathbf{n}_{\mathrm{RH}}(s)$ of them as they are the ones that are under the least interference. We refer to these chosen channels as the set $\mathcal{Q}^{*}_{s}$. In case of $\mathbf{n}_{\mathrm{RH}}(s) \geq Q_{s}$, all channels will be chosen; thus, we update the lookup table such that for $q \in \mathcal{Q}^{*}_{s}$, $LT(\mathbf{n}, s, q) = 1$. We continue the aforesaid process, from highest to lowest priority and once for each slice, until $LT$ for all slices is updated. This process has been summarized in Algorithm \ref{alg:lookup}.
\begin{algorithm}[!t]
	\caption{Lookup table construction}
	\label{alg:lookup}
	\begin{algorithmic}[1]
		\FOR{Each network state $\mathbf{n}$}
		\STATE  Set $LT(\mathbf{n}, s, q) = 0$ for all pairs $(s, q)$
		\STATE  Set slice $s$ priority based on $\mathbf{n}_{\mathrm{RH}}(s)/Q_s$
			\FOR{All $s\in\mathcal{S}$ and from highest priority to lowest}
			\STATE For $1\leq q \leq Q_{s}$ calculate $C_{s,q,u}(\mathbf{\Delta}_{s,q})$ using \eqref{eq:174806} considering that all channels that have the value 1 in $LT$ are interfering
			\STATE Sort channels according to $C_{s,q,u}(\mathbf{\Delta}_{s,q})$, choose the first $\mathbf{n}_{\mathrm{RH}}(s)$ ones and call it $\mathcal{Q}^{*}_{s}$
			\STATE For $q \in \mathcal{Q}^{*}_{s}$ set $LT(\mathbf{n}, s, q) = 1$
			\ENDFOR 
		\ENDFOR 
		\RETURN $LT$
	\end{algorithmic} 
\end{algorithm}

Having the $LT$ obtained from Algorithm \ref{alg:lookup}, one can use the state-dependent queueing model mentioned in Section \ref{sec:queueing}, with some modifications, to achieve the network steady-state probability for an interference-aware channel allocation policy. For each state $\mathbf{n}$ and slice $s$, the value $LT(\mathbf{n}, s, q)$ determines whether channel $q$ can be used or not. In state $\mathbf{n}$, total number of $\mathbf{n}_{\mathrm{RH}}(s)$ channels are allocated to slice $s$. We assume that a tagged flow can randomly be assigned to any of the allocated channels, and therefore, $\mathrm{Pr}(q)$ in \eqref{eq:669958} will be
\begin{equation} 
	\label{eq:555735}
	\mathrm{Pr}(q) =
	\begin{cases}
	\frac{1}{\min(\mathbf{n}_{\mathrm{RH}}(s),Q_s)},  &  \;\; LT(\mathbf{n}, s, q) = 1 \\
	0, & \;\; \text{o.w.}
	\end{cases}
\end{equation}
where $\min(\mathbf{n}_{\mathrm{RH}}(s),Q_s)$, shows the number of busy channels in slice $s$ in state $\mathbf{n}$. To be specific, in an interference-aware channel allocation policy, the channels to be allocated are known for each state, but since each slice supports a single type of service and therefore all flows that are allocated to a slice have the same priority, when calculating the average rate for a tagged flow, the assigned channel to that flow is random.

Furthermore, since allocated channels in state $\mathbf{n}$ and for all slices are known, the interference on channel $q$ of slice $s$ is also known and accordingly the interference vector is known, i.e., equal to $\mathbf{\Delta}^{*}_{s,q} \in \{0,1\}^{|\mathcal{N}_{s,q}| - 1}$. Therefore, \eqref{eq:990352} would be modified as
\begin{equation} 
	\label{eq:993352}
	\mathrm{Pr}(\mathbf{\Delta}_{s,q})= 
	\begin{cases}
		1, & \;\; \mathbf{\Delta}_{s,q} =\mathbf{\Delta}^{*}_{s,q} \\
		0, & \;\; \mathbf{\Delta}_{s,q} \neq \mathbf{\Delta}^{*}_{s,q} \\
	\end{cases}	.
\end{equation}

As we already mentioned, this channel allocation policy requires a cooperation among MVNOs. They also need to always know the number of customers $\mathbf{n}_{\mathrm{RH}}(s)$ and lookup table $LT$. Exchanging the lookup table happens only once at the beginning, but the number of customers has to be exchanged continually and therefore adds a data overhead to the network. Since $\mathbf{n}_{\mathrm{RH}}(s)$ is only a vector of non-negative integers, when its dimension is small, this overhead is negligible.

%% file: numerical.tex
\section{Numerical Results}  \label{sec:numerical}
In this section, first, we present numerical results for a single-MVNO single-SP scenario with symmetric slicing configuration to show the accuracy of our analytic model and the proposed approximation in Algorithm \ref{alg:beta} in comparison to discrete-event simulations (DES). Then, we consider a more comprehensive multi-MVNO multi-SP scenario with an asymmetric slicing configuration and examine the network KPIs and isolation metrics to evaluate the proposed interference-aware channel allocation policy compared to the random allocation policy and also the exhaustive search-based optimal policy as a benchmark. We use \textsc{Matlab} environment to achieve both analytic and DES results.

\subsection{Single-MVNO Single-SP Scenario} \label{198458}
In this part and in order to show the validity of our proposed model, we turn our attention to queues themselves with consideration of a single MVNO $v \in \mathcal{V} = \{1\}$ and a single SP $u \in \mathcal{U} = \{1\}$ and then extract slice KPIs. 

Similar to Fig. \ref{fig:Cell} and without loss of generality, we consider a cellular network with three BSs $b \in \mathcal{B} = \{1, 2, 3\}$ located at the center of identical hexagonal cells with cell radius, i.e., hexagon edge, equal to \SI{200}{\meter} and equal transmission powers. SP $u$-users that are uniformly distributed throughout the cells, initiate downloading data flows with its average size $\Omega_u = \SI{80}{Mbit}$ at random times following a Poisson point process. SP $u=1$ delivers all its flows to MVNO $v=1$ and therefore $P_{u,v}^{\mathrm{SP}} = 1$.

\begin{figure}[!t] 
	\centering \includegraphics[width=0.75\columnwidth]{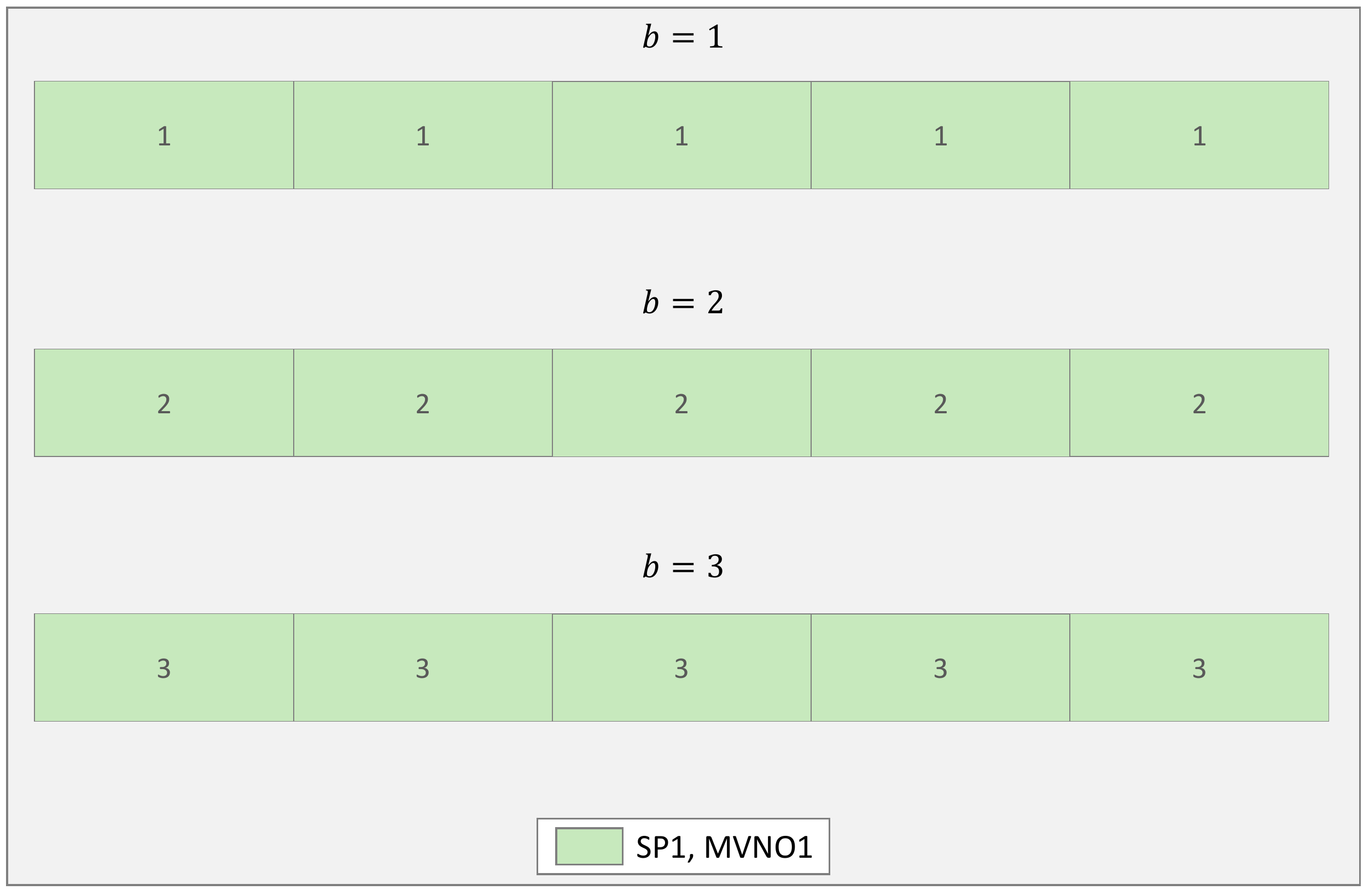}
	\caption{Single-MVNO single-SP slicing configuration}
	\label{fig:scenraio1}
\end{figure}

MVNO $v$ has only one slice in each cell and therefore $s \in \mathcal{S} = \{1, 2, 3\}$. We consider a symmetric slicing configuration as depicted in Fig \ref{fig:scenraio1} with $Q_{s}^{\mathrm{max}} = 10$ and $Q_{s} = 5$ channels with bandwidth $w_{s}=\SI{20}{\MHz}$ per channel for all $s \in \mathcal{S}$. BS transmission power $P^{\mathrm{BS}}_{b}$ in each cell is uniformly distributed throughout the specified bandwidth. MVNO $v$, in each cell, assigns all flows from SP $u$ to its slice and therefore, for all $s\in \mathcal{S}$ we have $P_{u,v,s}^{\mathrm{MVNO}} = 1$ and also as cells are similar, we have $\lambda_{s,u} = \lambda_{u}/3$. Additional parameters are as in Table \ref{table:sim}.

\begin{table}[!t] 
	\centering
	\caption{Numerical parameters}
	\label{table:sim}
	\begin{tabular}{lr}
		\hline
		\multicolumn{2}{c}{\textbf{Common configurations}}\\
		Path loss model & $128.1 + 37.6 log10(d/\SI{}{\km})$ \SI{}{\dB} \\
		Carrier frequency &  \SI{2.0}{\GHz}\\
		Cell radius &  \SI{200}{\meter}\\
		Thermal noise &  \SI{-174}{dBm/\Hz}\\
		Bandwidth efficiency $\eta_1$& 0.63 \\
		SINR efficiency $\eta_2$& 0.4 \\
		Number of packets generated in DES & 300000 \\
		Number of iterations for Algorithm \ref{alg:beta} & 20 \\
		Power Bandwidth& \SI{72}{\MHz} \\
		\hline
		\multicolumn{2}{c}{\textbf{Single MVNO - Single SP configurations}}\\
		Average number of bits $\Omega_{u}$  & \SI{80}{Mbit}\\
		Slice channel bandwidth $w_{s}$  & \SI{20}{\MHz}\\
		Number of slice channels $Q_{s}$ & 5\\
		Queue capacity $Q_{s}^{\mathrm{max}}$ & 10\\
		\hline
		\multicolumn{2}{c}{\textbf{Multi MVNO - Multi SP configurations}}\\
		Maximum transmission power $P^{\mathrm{BS}}_{b}$ & \SI{45}{dBm}\\
		SP 1-users channel bandwidth $w_{s}$  & \SI{18}{\MHz}\\
		SP 2-users channel bandwidth $w_{s}$  & \SI{6}{\MHz}\\
		Average number of bits $\Omega_{1}$  & \SI{8}{Mbit}\\
		Average number of bits $\Omega_{2}$  & \SI{80}{Mbit}\\
		\hline
	\end{tabular}
\end{table}  

To evaluate the accuracy of our state-dependent approach, we consider DES and compare delay and throughput results. We also compare our work with the averaged interference method proposed in \cite{bib:958292}. In averaged interference method, an iterative approach is used to calculate the network steady-state probability distribution. Starting with an initial value for the network steady-state probability distribution in the first iteration, the denominator of SINR in \eqref{eq:919125}, is replaced by an equivalent interfering power that obtained by averaging over interfering powers from neighboring cells with network steady-state probabilities as weights. Afterwards, the equivalent service rate for each queue is achieved. Then, the network is decomposed into individual queues and using the obtained equivalent service rate, the steady-state probability distribution is calculated independently which is used in the next iteration until convergence.

\begin{figure*} 
	\centering \includegraphics[width=\textwidth,trim={3cm 0cm 2.75cm 1cm}, clip=true]{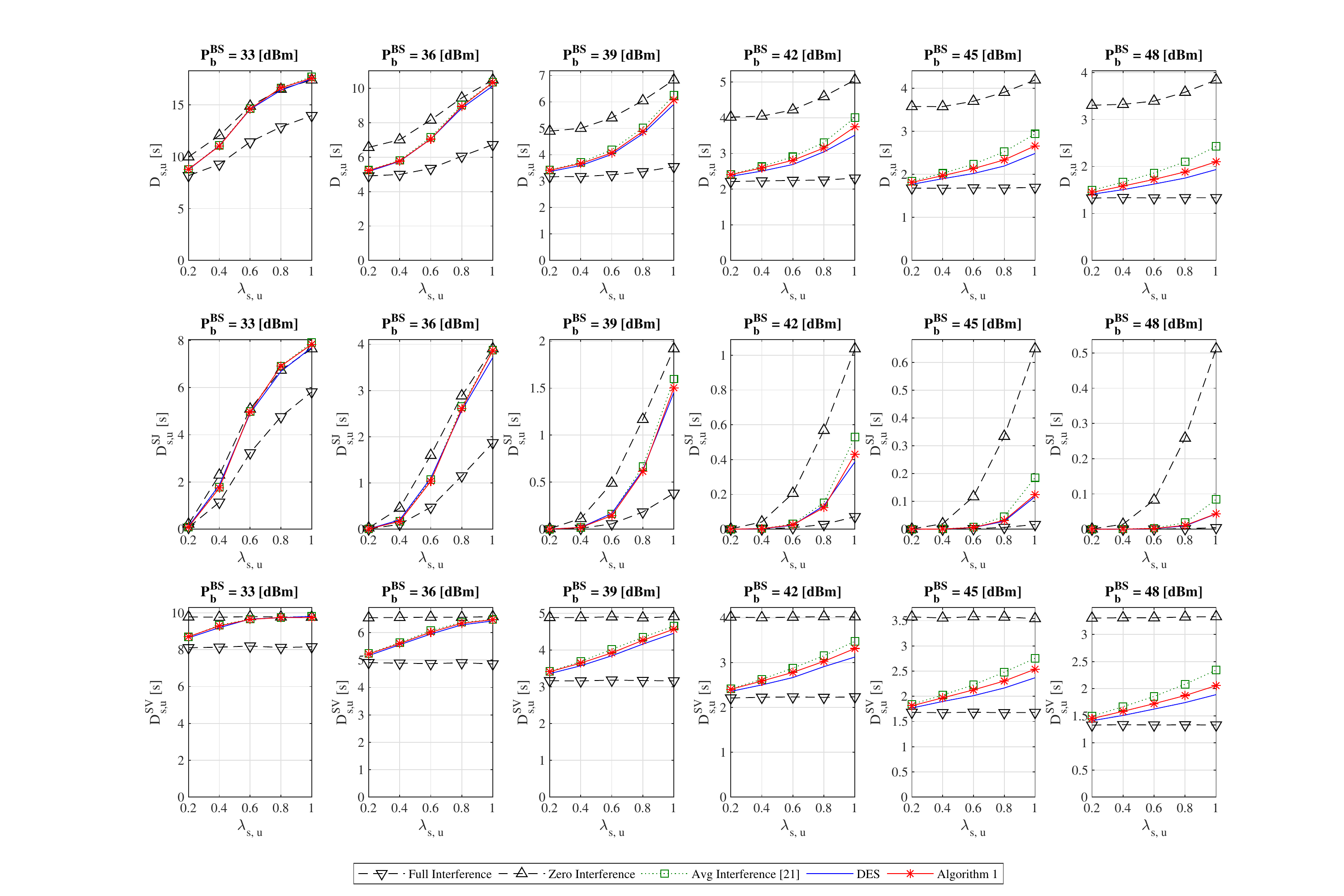}
	\caption{$D_{s, u}$, $D^{\mathrm{SJ}}_{s, u}$ and $D^{\mathrm{SV}}_{s, u}$ for the typical slice $s\in \mathcal{S}$ versus BS power and flow traffic when considering a random channel allocation.}
	\label{fig:Delay_random}
\end{figure*}

\begin{figure*} 
	\centering \includegraphics[width=\textwidth,trim={2.7cm 7cm 2.7cm 0.5cm}, clip=true]{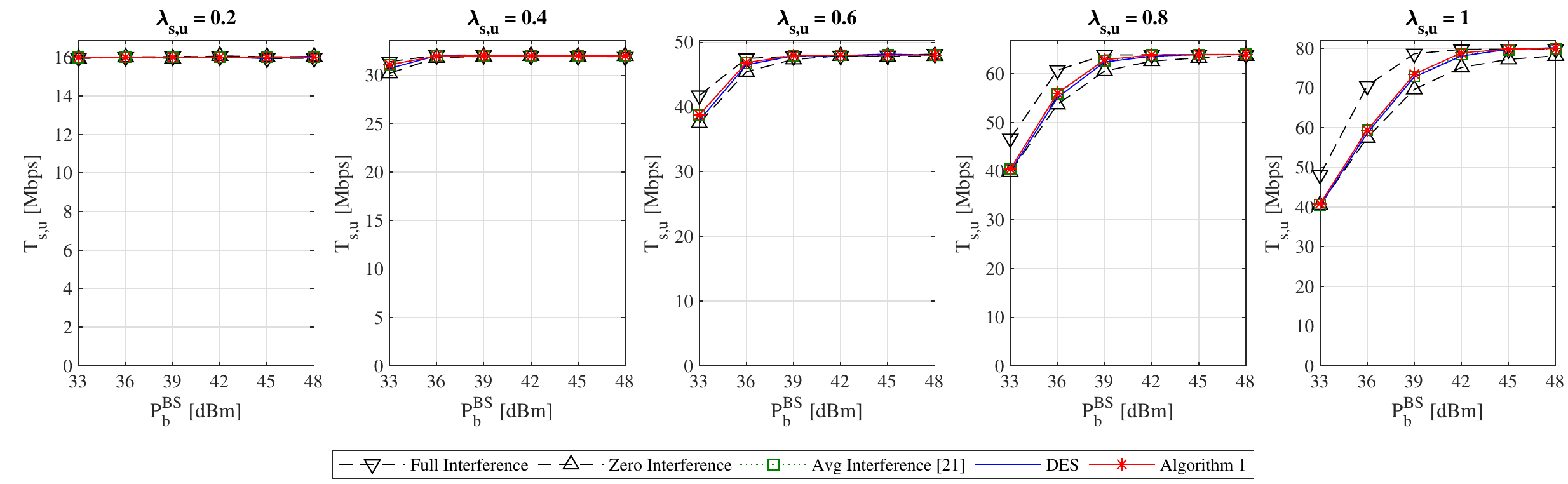}
	\caption{$T_{s, u}$ for the typical slice $s\in \mathcal{S}$ versus BS power and flow traffic in neighboring cells when considering a random channel allocation.}
	\label{fig:tp_random}
\end{figure*}

We test the network for multiple values of SP $u$-users traffic $\lambda_{u}$ and maximum transmission power $P^{\mathrm{BS}}_{b}$. Fig. \ref{fig:Delay_random} shows $D_{s, u}$, $D^{\mathrm{SJ}}_{s, u}$ and $D^{\mathrm{SV}}_{s, u}$ for the typical slice $s\in \mathcal{S}$ when using a random channel allocation. When maximum transmission power has the lowest value, i.e., $P^{\mathrm{BS}}_{b} = \SI{33}{dBm}$, average difference between DES and Algorithm \ref{alg:beta} for $D_{s, u}$ is $0.5\%$. As $P^{\mathrm{BS}}_{b}$ increases from $\SI{33}{dBm}$ to $\SI{48}{dBm}$, the average difference increases from $0.5\%$ to $6.0\%$. Meanwhile, the average difference between DES and the averaged interference method, increases from $0.7\%$ to $15.3\%$. Furthermore, Fig. \ref{fig:tp_random} shows $T_{s, u}$ for the typical slice $s\in \mathcal{S}$ and it can be seen that increasing $P^{\mathrm{BS}}_{b}$ hardly affects the average difference between DES, Algorithm \ref{alg:beta} and averaged interference method which is about $0.5\%$.

Based on the results, one can conclude that the proposed state-dependent queueing network model is more accurate than the other method. The justification for this accuracy is that, in our proposed method, the service rate changes proportional to the interference changes. But in the averaged interference method, an equivalent averaged service rate represents all interfering states. Hence, when the difference between the full and zero interference states is large, an averaged service rate is not sufficiently accurate in modeling interference-coupled slices. Despite this, the accuracy of our proposed method is limited to the approximation in Algorithm \ref{alg:beta} and the cost function \eqref{eq:214038}. 

Next, we try to evaluate the efficiency of the interference-aware channel allocation. First, using Algorithm \ref{alg:lookup}, MVNO obtains an interference-aware lookup table $LT$ that dictates which channels must be allocated to each slice. We use the same $LT$ in both DES and state-dependent queueing network model. For the latter, we use $LT$ along with Algorithm \ref{alg:beta} as explained in Section \ref{sec:allocation}. In DES, each time the network state changes, MVNO assigns channels based on the $LT$ values. To be in accordance with the analysis, once the allocated channels for a slice are known, we assume flows that belong to that slice are randomly assigned to the allocated channels. Therefore, the channel allocation is deterministic, but flow assignment to allocated channels within a slice still happens randomly similar to the analysis. 

\begin{figure*} 
	\centering \includegraphics[width=\textwidth,trim={3cm 0cm 2.75cm 1cm}, clip=true]{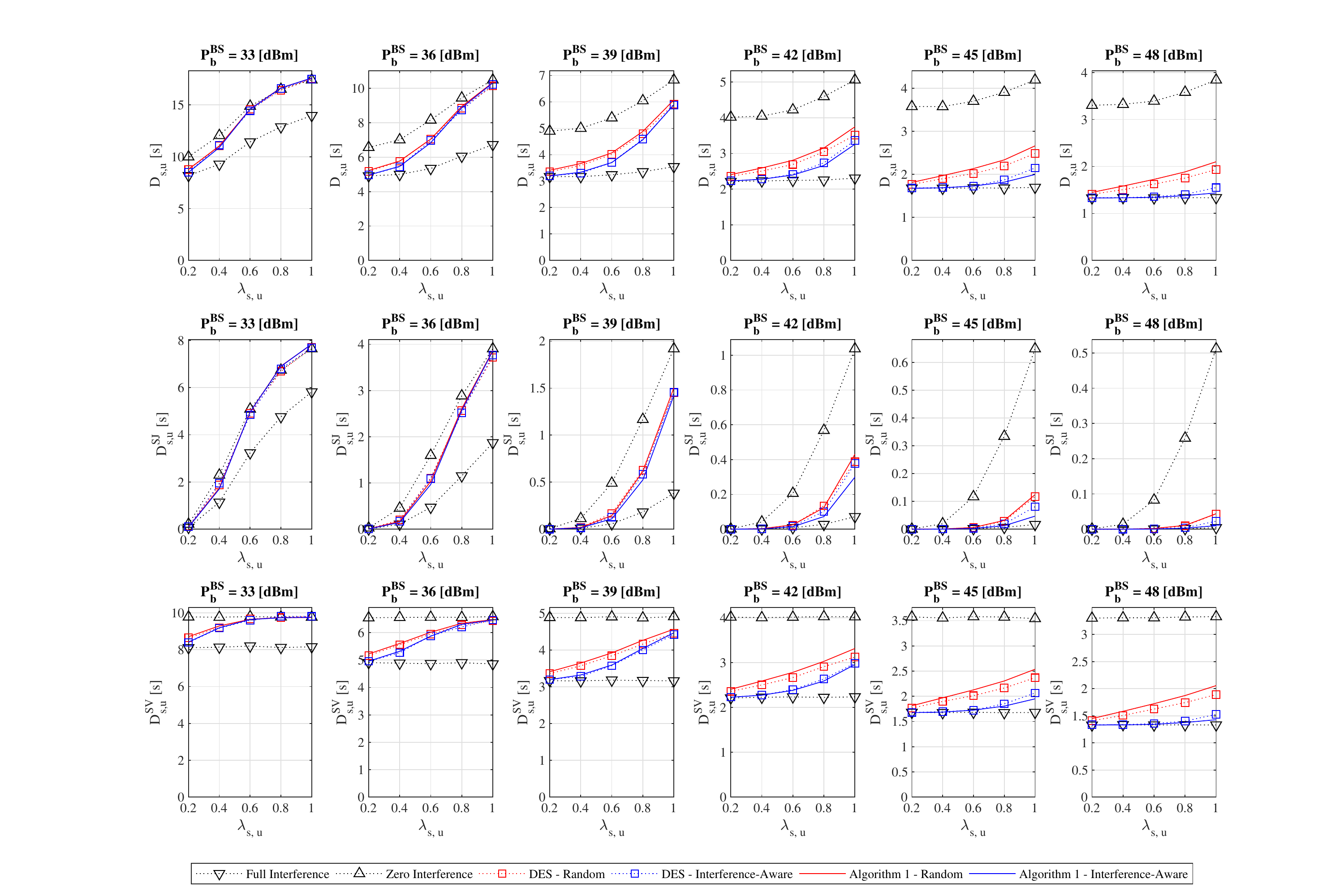}
	\caption{$D_{s, u}$, $D^{\mathrm{SJ}}_{s, u}$ and $D^{\mathrm{SV}}_{s, u}$ for the typical slice $s\in \mathcal{S}$ versus BS power and flow traffic when considering an interference-aware channel allocation.}
	\label{fig:Delay_smart}
\end{figure*}

Fig. \ref{fig:Delay_smart} shows $D_{s, u}$, $D^{\mathrm{SJ}}_{s, u}$ and $D^{\mathrm{SV}}_{s, u}$ for a typical slice $s\in \mathcal{S}$. The average total delay $D_{s, u}$ for interference-aware channel allocation is on average $0.9\%$ less than the random channel allocation when $P^{\mathrm{BS}}_{b} = \SI{33}{dBm}$. As the maximum transmission power increases from $\SI{33}{dBm}$ to $\SI{48}{dBm}$, i.e., when the interference is intensified, the average difference increases from $0.9\%$ to $20.9\%$. Therefore, the interference-aware channel allocation can effectively reduce the interference. Fig. \ref{fig:Delay_smart} also reveals that, when flow traffic is extremely low or high, interference-aware and random channel allocation almost have similar performance. The reason is, when the traffic is extremely low, the effect of interference is negligible. However, when the number of customers is moderate, Algorithm \ref{alg:lookup} can effectively arrange allocations and increase the service rate. But, as the number of customers increases, interference-aware allocation loses its efficiency because sufficient flexibility in channel allocation does not exist. 

\begin{figure*}  
	\centering \includegraphics[width=\textwidth,trim={2.7cm 7cm 2.7cm 0.5cm}, clip=true]{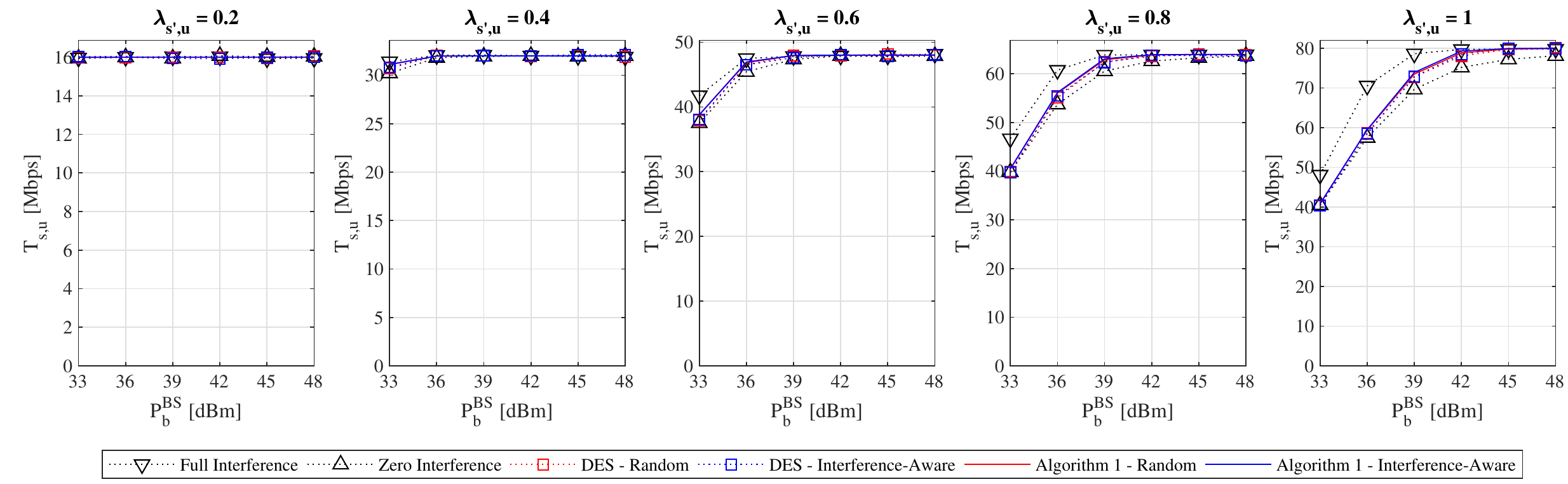}
	\caption{$T_{s, u}$ for the typical slice $s\in \mathcal{S}$ versus BS power and flow traffic in neighboring cells when considering an interference-aware channel allocation.}
	\label{fig:tp_smart}
\end{figure*}

Furthermore, Fig. \ref{fig:tp_smart} shows $T_{s, u}$ for the typical slice $s\in \mathcal{S}$. It is apparent that the difference between interference-aware and random channel allocation for $T_{s, u}$ is minute and interference-aware policy on average increases the throughput by $0.1\%$. This is because the interference hardly affects the throughput as we did not consider any expiration age for the arrived packets. Fig. \ref{fig:Delay_smart} also shows that, when using the interference-aware policy and maximum transmission power has the lowest value, i.e., $P^{\mathrm{BS}}_{b} = \SI{33}{dBm}$, average difference between DES and Algorithm \ref{alg:beta} for $D_{s, u}$ is $0.8\%$. As $P^{\mathrm{BS}}_{b}$ increases from $\SI{33}{dBm}$ to $\SI{48}{dBm}$, the average difference increases from $0.8\%$ to $2.2\%$. Besides, Fig. \ref{fig:tp_smart} shows that the average difference between DES and Algorithm \ref{alg:beta} for $T_{s, u}$ is small and equal to $0.6\%$. 

\subsection{Multi-MVNO Multi-SP Scenario}
In this part and in order to investigate the inter-slice isolation, we consider a more comprehensive scenario with asymmetric slicing configuration similar to Fig. \ref{fig:Slice} with $\mathcal{S} = \{1,\dots,12\}$, $\mathcal{V} = \{1, 2\}$ and $\mathcal{U} = \{1, 2\}$. For slices $s \in \{1,4,5,7,9,11\}$ channel bandwidth $w_{s} = \SI{6}{\MHz}$, the number of channels $Q_{s} = 3$ and there are no buffers, i.e.,  $Q^{\mathrm{max}}_s = 3$. For slices $s \in \{2,3,6,8,10,12\}$ channel bandwidth $w_{s} = \SI{18}{\MHz}$, the number of channels $Q_{s} = 1$ and we consider a single buffer, i.e.,  $Q^{\mathrm{max}}_s = 2$.

We consider the same cellular network described in \ref{198458} with maximum transmission power $P^{\mathrm{BS}}_{b}=\SI{45}{dBm}$ for all cells $b \in \mathcal{B}$. SP $1$-users and SP $2$-users are distributed throughout the network coverage area with a two-dimensional uniform spatial distribution. We consider SP $1$ as an example of a delay-sensitive service (like URLLC service) with average flow size $\Omega_1 = \SI{8}{Mbit}$ and SP $2$ as an example of a throughput-sensitive service (like eMBB service) with average flow size $\Omega_2 = \SI{80}{Mbit}$. MVNO $1$ assigns flows of SP $1$-users to slices $s \in \{1,7,9\}$, i.e., $P_{1,1,s}^{\mathrm{MVNO}} = 1$ and flows of SP $2$-users to slices $s\in \{3,6,12\}$, i.e., $P_{2,1,s}^{\mathrm{MVNO}} = 1$. MVNO $2$ assigns flows of SP $1$-users to slices $s \in \{4,5,12\}$ , i.e., $P_{1,2,s}^{\mathrm{MVNO}} = 1$ and flows of SP $2$-users to slices $s\in \{2,8,10\}$, i.e., $P_{2,2,s}^{\mathrm{MVNO}} = 1$. For unmentioned cases, we have $P_{u,v,s}^{\mathrm{MVNO}} = 0$. Additional parameters are listed in Table \ref{table:sim}.

\begin{figure*}[h!] 
	\begin{subfigure}{\textwidth}
		\centering \includegraphics[width=\textwidth,trim={2cm 6.4cm 2cm 0.5cm}, clip=true]{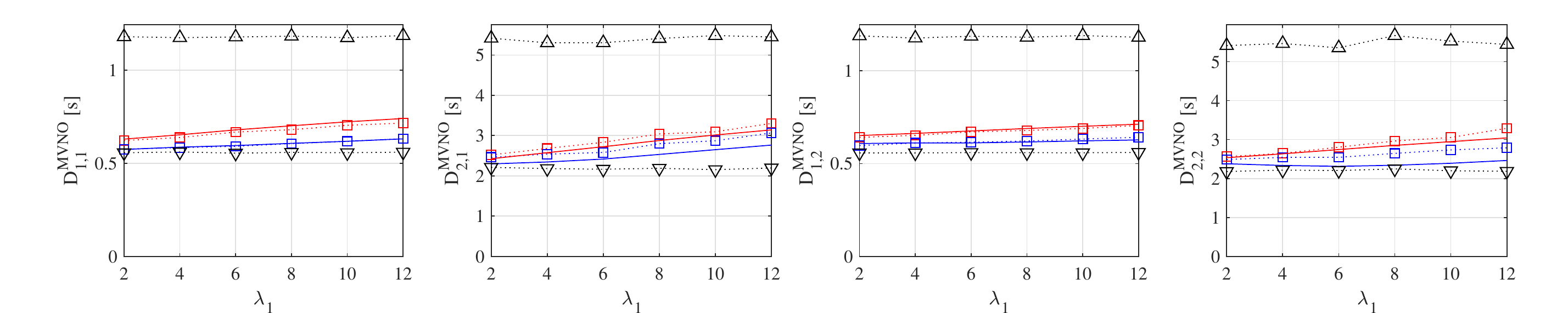}
		\caption{Increasing $\lambda_{1}$ when $\lambda_{2}=0.6$ flows per second}
		\label{fig:lambda_sweep_delay_a}
	\end{subfigure}
\hfill
	\begin{subfigure}{\textwidth}
		\centering \includegraphics[width=\textwidth,trim={2cm 5cm 2cm 0.7cm}, clip=true]{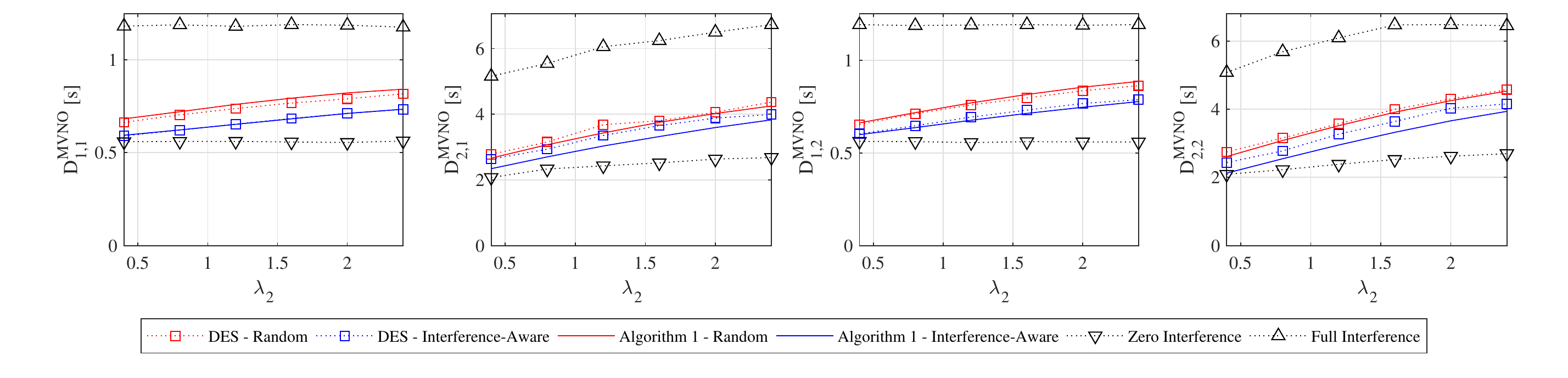}
		\caption{Increasing $\lambda_{2}$ when $\lambda_{1}=8$ flows per second}
		\label{fig:lambda_sweep_delay_b}
	\end{subfigure}
	\caption{Network total delay $D^{\mathrm{MVNO}}_{u, v}$ versus SP $u$-users flow traffic for the multi-MVNO multi-SP scenario.}
	\label{fig:lambda_sweep_delay}
\end{figure*}

\begin{figure*}[h!] 
	\begin{subfigure}{0.5\textwidth}
		\centering \includegraphics[width=\textwidth,trim={1cm 0cm 2cm 0cm}, clip=true]{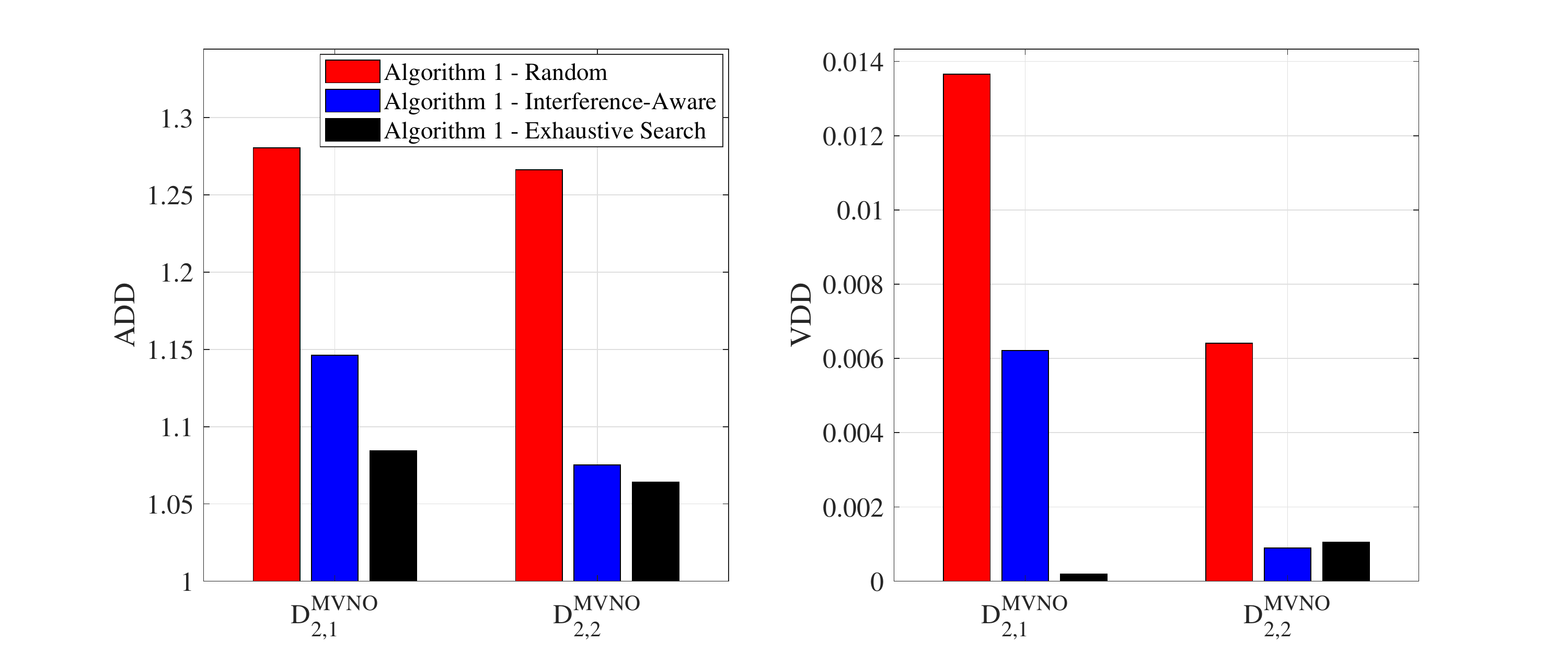}
		\caption{Increasing $\lambda_{1}$ when $\lambda_{2}=0.6$ flows per second}
		\label{fig:lambda_sweep_isolation_a}
	\end{subfigure}
	\begin{subfigure}{0.5\textwidth}
		\centering \includegraphics[width=\textwidth,trim={2cm 0cm 1cm 0cm}, clip=true]{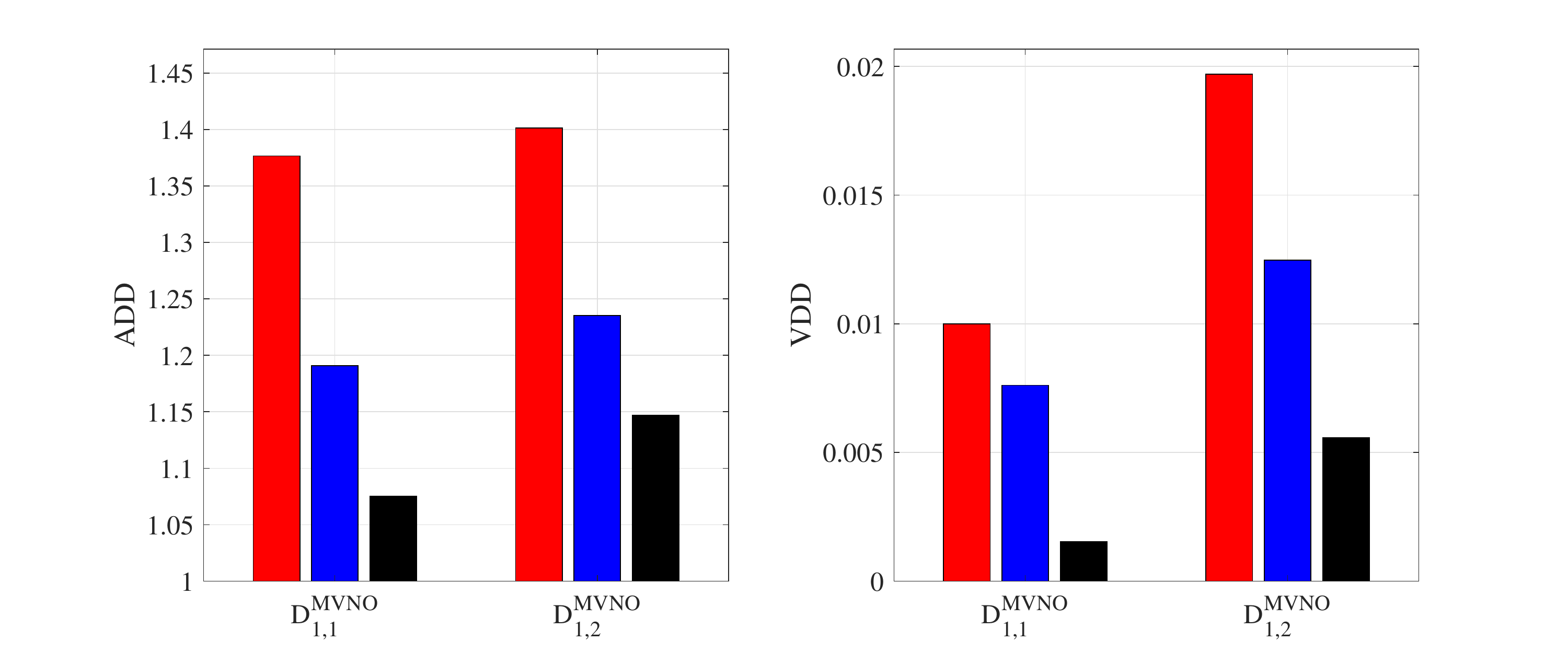}
		\caption{Increasing $\lambda_{2}$ when $\lambda_{1}=8$ flows per second}
		\label{fig:lambda_sweep_isolation_b}
	\end{subfigure}
	\caption{Average and variance of $D^{\mathrm{MVNO}}_{u, v}$ versus SP $u$-users flow traffic for the multi-MVNO multi-SP scenario.}
	\label{fig:lambda_sweep_isolation}
\end{figure*}

Each SP $u \in \mathcal{U}$ delivers its flows to MVNOs $v \in \mathcal{V}$ with equal probability and therefore $P_{u,v}^{\mathrm{SP}} = 0.5$. Since SP $2$-users are throughput-sensitive, to evaluate the effect of SP $1$-users activity on SP $2$-users throughput, we keep $\lambda_{2} = 0.6$ constant and sweep $\lambda_{1}$. Consequently, each MVNO carries \SI{24}{Mbps} traffic for SP $2$-users and \SI{8}{Mbps} to \SI{48}{Mbps} for SP $1$-users (regarding negligible blocking). Fig. \ref{fig:lambda_sweep_delay_a} shows the network total delay for all SPs and MVNOs. In the case of random channel allocation and when $\lambda_{1} = 2$, i.e., low SP $1$-users traffic, interference causes the SP $2$-users delay $D^{\mathrm{MVNO}}_{2, 1}$ and $D^{\mathrm{MVNO}}_{2, 2}$ to be increased by $13.1\%$ compared to zero interference condition. Since the interference is coupled, SP $1$-users delay $D^{\mathrm{MVNO}}_{1, 1}$ and $D^{\mathrm{MVNO}}_{1, 2}$ also increase by $14.7\%$ compared to zero interference condition. The increase in SP $1$-users traffic up to $500\%$, increases SP $2$-users delay $D^{\mathrm{MVNO}}_{2, 1}$ and $D^{\mathrm{MVNO}}_{2, 2}$ up to $24.8\%$ compared to low SP $1$-users traffic condition. Using an interference-aware channel allocation can reduce $D^{\mathrm{MVNO}}_{2, 1}$ and $D^{\mathrm{MVNO}}_{2, 2}$, respectively, by $12.0\%$ and $19.0\%$ in case of highest SP $1$-users traffic compared to random channel allocation. The reason for the difference in KPI improvement between MVNO $1$ and MVNO $2$ is that in the latter, slices that are allocated to SP $2$-users, have less overlap with all slices that are allocated to SP $1$-users. This improvement has also been reflected in Fig. \ref{fig:lambda_sweep_isolation_a} where both ADD and VDD have been decreased for interference-aware channel allocation and MVNO $2$ slices are more isolated. Furthermore, interference-aware channel allocation also reduces $D^{\mathrm{MVNO}}_{1, 1}$ and $D^{\mathrm{MVNO}}_{1, 2}$ by $13.5\%$ in case of highest SP $1$-users traffic and compared to random channel allocation by mitigating the coupled interference more.

\begin{figure*}[h!] 
	\begin{subfigure}{\textwidth}
		\centering \includegraphics[width=\textwidth,trim={2cm 6.4cm 2cm 0.5cm}, clip=true]{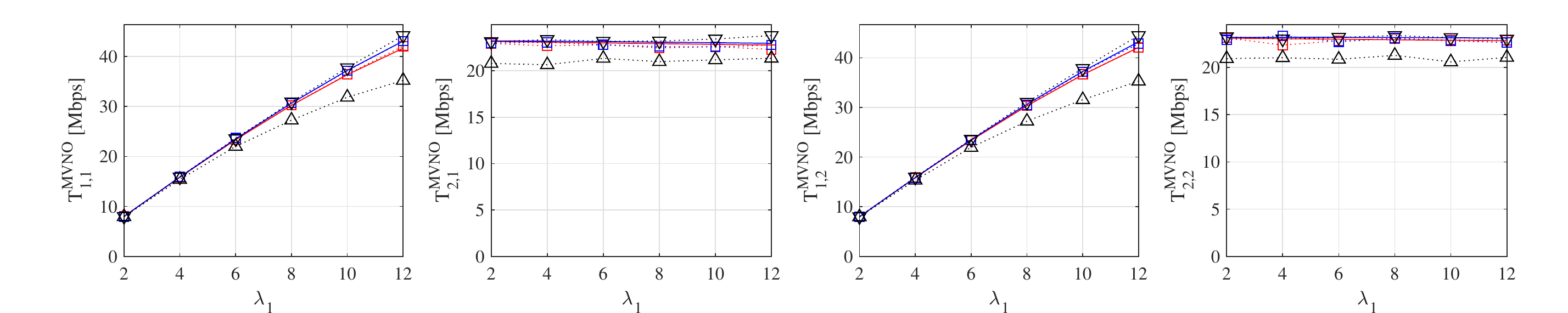}
		\caption{Increasing $\lambda_{1}$ when $\lambda_{2}=0.6$ flows per second}
		\label{fig:lambda_sweep_tp_a}
	\end{subfigure}
	\hfill
	\begin{subfigure}{\textwidth}
		\centering \includegraphics[width=\textwidth,trim={2cm 5cm 2cm 0.7cm}, clip=true]{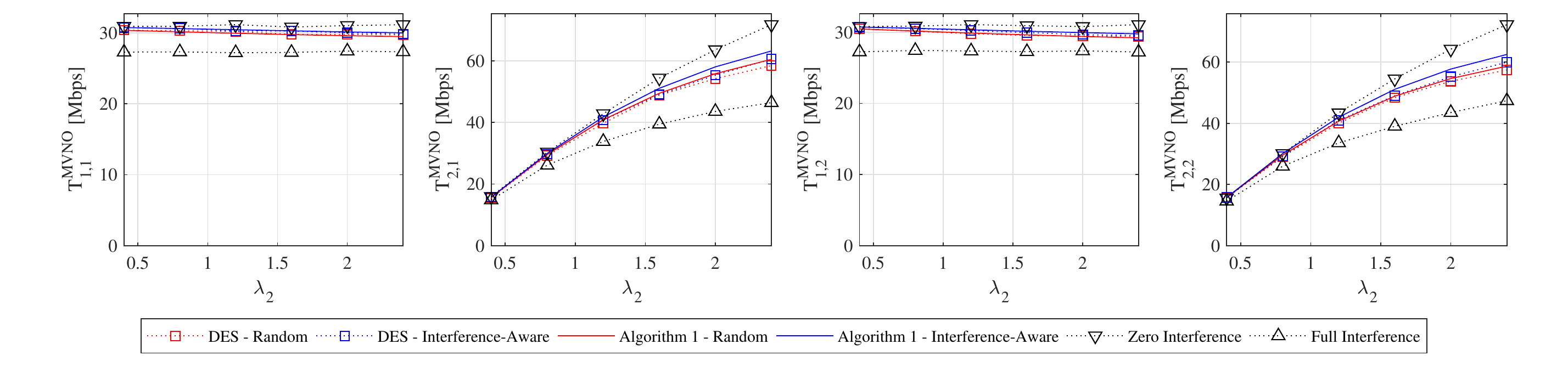}
		\caption{Increasing $\lambda_{2}$ when $\lambda_{1}=8$ flows per second}
		\label{fig:lambda_sweep_tp_b}
	\end{subfigure}
	\caption{$T^{\mathrm{MVNO}}_{u, v}$ versus SP $u$-users flow traffic for the multi-MVNO multi-SP scenario.}
	\label{fig:lambda_sweep_tp}
\end{figure*}

\begin{figure*}[h!] 
	\begin{subfigure}{0.5\textwidth}
		\centering \includegraphics[width=\textwidth,trim={1cm 0cm 2cm 0cm}, clip=true]{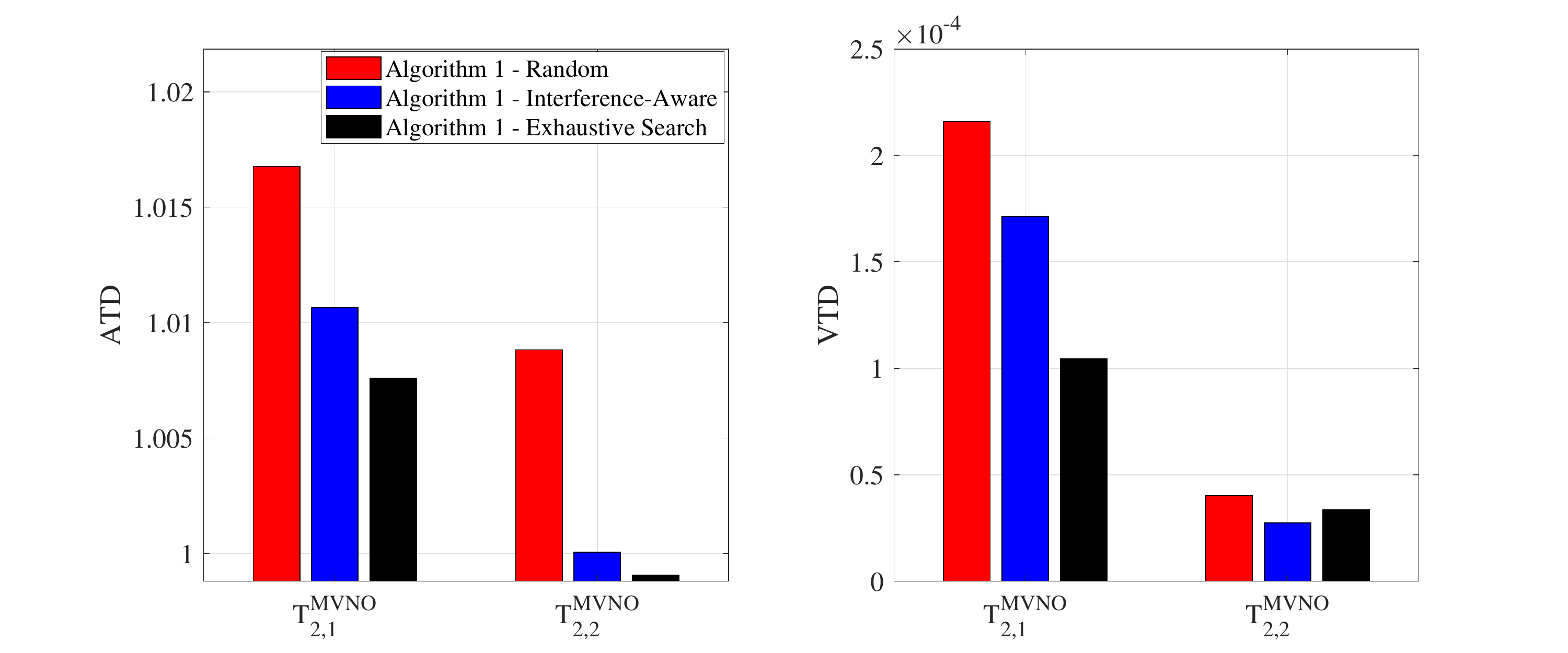}
		\caption{Increasing $\lambda_{1}$ when $\lambda_{2}=0.6$ flows per second}
		\label{fig:lambda_sweep_isolation_tp_a}
	\end{subfigure}
	\begin{subfigure}{0.5\textwidth}
		\centering \includegraphics[width=\textwidth,trim={1.9cm 0cm 1cm 0cm}, clip=true]{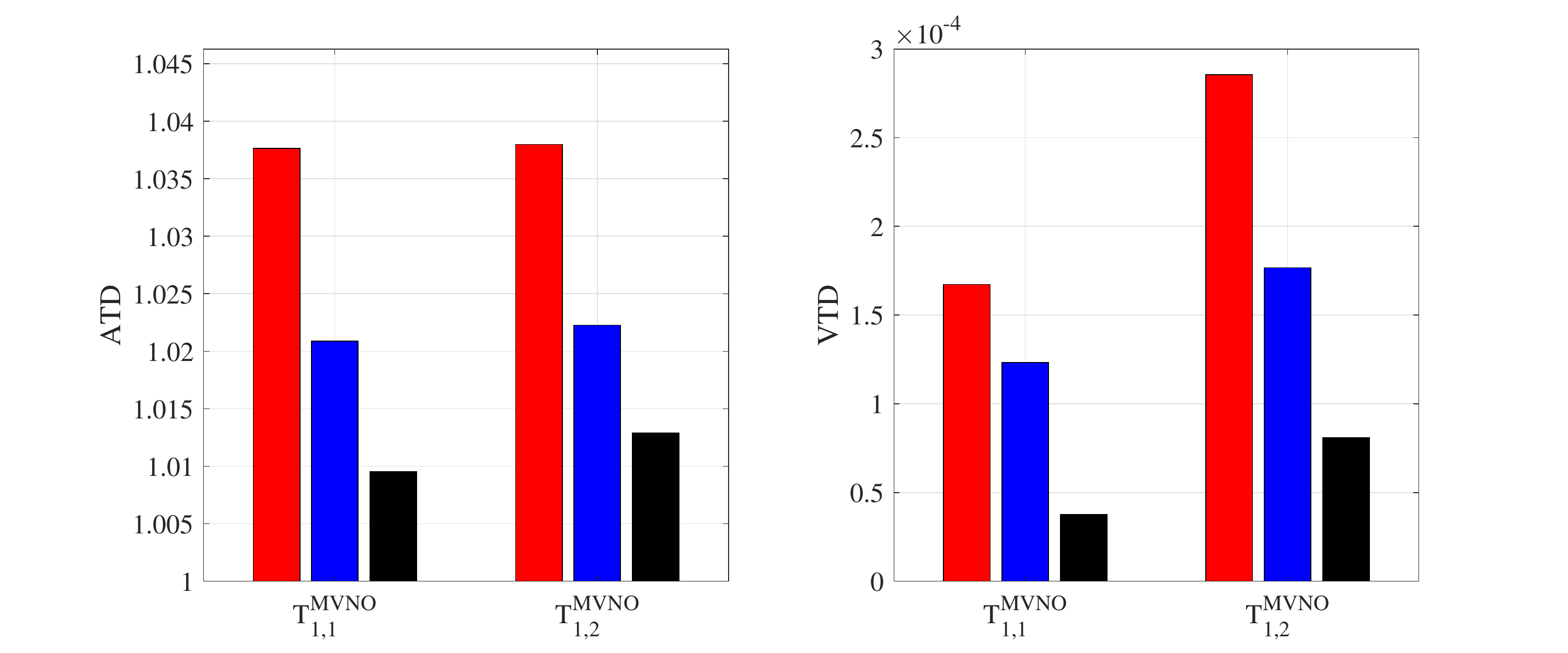}
		\caption{Increasing $\lambda_{2}$ when $\lambda_{1}=8$ flows per second}
		\label{fig:lambda_sweep_isolation_tp_b}
	\end{subfigure}
	\caption{Average and variance of $T^{\mathrm{MVNO}}_{u, v}$ versus SP $u$-users flow traffic for the multi-MVNO multi-SP scenario.}
	\label{fig:lambda_sweep_isolation_tp}
\end{figure*}

 Fig. \ref{fig:lambda_sweep_tp_a} shows that the network throughput $T^{\mathrm{MVNO}}_{2, 1}$ and $T^{\mathrm{MVNO}}_{2, 2}$ are decreased by $0.5\%$ compared to zero interference condition in case of low SP $1$-users traffic. Increasing SP $1$-users traffic up to $500\%$ results in a decrease in $T^{\mathrm{MVNO}}_{2, 1}$ and $T^{\mathrm{MVNO}}_{2, 2}$ by $1.7\%$ and $1.2\%$ compared to low SP $1$-users traffic condition. Using an interference-aware channel allocation increases the $T^{\mathrm{MVNO}}_{2, 1}$ and $T^{\mathrm{MVNO}}_{2, 2}$ by $0.8\%$ and $1.2\%$ in case of highest SP $1$-users traffic and compared to random channel allocation. This slight improvement is also evident in Fig. \ref{fig:lambda_sweep_isolation_tp_a} where both ATD and VTD have been decreased for interference-aware channel allocation.
 
Next, to evaluate the effect of SP $2$-users activity on SP $1$-users delay, we keep $\lambda_{1} = 8$ and sweep $\lambda_{2}$. Consequently, each MVNO carries \SI{32}{Mbps} traffic for SP $1$-users and \SI{16}{Mbps} to \SI{96}{Mbps} for SP $2$-users. Fig. \ref{fig:lambda_sweep_delay_b} shows that in the case of random channel allocation and when $\lambda_{2} = 0.4$, i.e., low SP $2$-users traffic, the network total delay $D^{\mathrm{MVNO}}_{1, 1}$ and $D^{\mathrm{MVNO}}_{1, 2}$ increases by $20.1\%$ compared to zero interference condition. The increase in SP $2$-users traffic up to $500\%$, increases SP $1$-users delay $D^{\mathrm{MVNO}}_{1, 1}$ and $D^{\mathrm{MVNO}}_{1, 2}$, respectively, up to $23.4\%$ and $33.8\%$ compared to low SP $2$-users traffic condition. Using the interference-aware channel allocation can reduce $D^{\mathrm{MVNO}}_{1, 1}$ and $D^{\mathrm{MVNO}}_{1, 2}$, respectively, by $12.8\%$ and $12.3\%$ in case of highest SP $2$-users traffic compared to random channel allocation. The differences in MVNO $1$ and $2$ values are due to the fact that MVNO $1$ slices that are allocated to SP $1$-users, have less overlap with all slices that are allocated to SP $2$-users. This improvement also has been reflected in Fig. \ref{fig:lambda_sweep_isolation_b} where both ADD and VDD have been decreased for interference-aware channel allocation, then MVNO $1$ slices are more isolated.

 Fig. \ref{fig:lambda_sweep_tp_b} shows that the network throughput $T^{\mathrm{MVNO}}_{1, 1}$ and $T^{\mathrm{MVNO}}_{1, 2}$ are decreased by $1.8\%$ compared to zero interference condition in case of low SP $2$-users traffic. Increasing SP $2$-users traffic up to $500\%$ results in a decrease in $T^{\mathrm{MVNO}}_{1, 1}$ and $T^{\mathrm{MVNO}}_{1, 2}$ by $2.9\%$ and $4.1\%$ compared to low SP $2$-users traffic condition. Using the interference-aware channel allocation increases $T^{\mathrm{MVNO}}_{1, 1}$ and $T^{\mathrm{MVNO}}_{1, 2}$ by $1.8\%$ in case of highest SP $2$-users traffic and compared to random channel allocation. This improvement is also evident in Fig. \ref{fig:lambda_sweep_isolation_tp_b} where both ATD and VTD have been decreased for interference-aware channel allocation. 

From what mentioned above and by comparing average and variance values in Figs. \ref{fig:lambda_sweep_isolation} and \ref{fig:lambda_sweep_isolation_tp}, it can be concluded that delay is generally more affected by interference than throughput whether the service under investigation is delay-sensitive or throughput-sensitive. This is because we did not consider any expiration age for arrived packets. Consequently, delay-sensitive services are more prone to breaching the SLA compared with throughput-sensitive services. The proposed interference-aware channel allocation can improve this problem by bringing more inter-slice isolation. To establish a better understanding, as a benchmark, we also compare the average and variance values to the results the optimal allocation policy obtained by an exhaustive search for constructing the lookup table. The improvement in results compared to the proposed interference-aware channel allocation policy is obvious. However, if we define complexity as the number of times that $C_{s,q,u}(\mathbf{\Delta}_{s,q})$ needs to be calculated during the construction of the lookup table, then the complexity of the proposed policy is $\sum_{\mathbf{n}} \sum_{s=1}^{|\mathcal{S}|} \text{sgn}(\mathbf{n}_{\mathrm{RH}}(s))Q_{s}$, as for each state $\mathbf{n}$ and for all channels of all slices that have at least one customer, the value of $C_{s,q,u}(\mathbf{\Delta}_{s,q})$ needs to be calculated once. On the other hand, the complexity of the exhaustive search is $\sum_{\mathbf{n}} \sum_{s=1}^{|\mathcal{S}|} \min{(\mathbf{n}_{\mathrm{RH}}(s), Q_{s})} \prod_{s=1}^{|\mathcal{S}|}\binom{Q_{s}}{\min{(\mathbf{n}_{\mathrm{RH}}(s), Q_{s})}}$, as for each state $\mathbf{n}$, the value of $C_{s,q,u}(\mathbf{\Delta}_{s,q})$ needs to be calculated for all permutations of busy channels which is significantly more. In our scenario, the complexity of the proposed policy is $5.22\times 10^7$ while the complexity of the exhaustive search is $2.86\times 10^9$.

%% file: conclusion.tex
\section{Conclusion} \label{sec:conclusion}
In this work, we investigated the isolation between slices in neighboring cells in a multi-cell RAN slicing scenario. We first explained that the main bottleneck in the inter-slice isolation problem is the coupled-interference in multi-cell networks. We proposed an analytical model comprised of  a network of state-dependent queues. Then, to achieve the network steady-state probability distribution, we proposed an iterative algorithm. Further, we defined KPIs that quantify QoS and isolation. At last, we proposed and analyzed an interference-aware channel allocation policy aiming to improve the defined KPIs by improving isolation between slices. Our numerical results demonstrated that, the proposed queueing theoretic model fairly follows the simulation results in terms of slice delay and throughput. Besides, our proposed interference-aware policy improved our defined isolation metrics compared to a random allocation policy.

%% file: appendix.tex
\appendices
\clearpage
\section{Proof for Proposition \ref{prop:111737}}
\label{app1}
Let us consider a departure from slice $s$ when the network is in state $\mathbf{n}$. From Theorem \ref{theory:951737} we have
\begin{equation} \nonumber
	\label{eq:755126}
	\xi(\mathbf{n}, \mathbf{a}) = \frac{\Psi (\mathbf{n} - \mathbf{a})}{\Phi (\mathbf{n})} = \frac{\Psi_{1} (\mathbf{n} - \mathbf{a}) \Psi_{2} (\mathbf{n} - \mathbf{a})}{\Phi_{1} (\mathbf{n})\Phi_{2} (\mathbf{n})}.
\end{equation}
From \eqref{eq:300065} we have 
\begin{equation} \nonumber
\Psi_{1} (\mathbf{n} - \mathbf{a}) \;=\; \Phi_{1} (\mathbf{n} - \mathbf{a} + \mathbf{e}(\mathbf{n} - \mathbf{a})) \prod_{s'=1}^{|\mathcal{S}|}\Big(1 - (1 - \Lambda_{s',u_{s'}}(Q^{\mathrm{max}}_{s'}))\delta(1 + \mathbf{n}_{\mathrm{LH}}(s') - \mathbf{a}_{\mathrm{LH}}(s')).
\end{equation}
For a departure movement from slice $s$ we have $\mathbf{a}_{\mathrm{RH}}(s) =1$ while all other elements of $\mathbf{a}$ are equal to zero. Therefore, we have $\delta(1 + \mathbf{n}_{\mathrm{LH}}(s')-\mathbf{a}_{\mathrm{LH}}(s')) =0$ as $\mathbf{n}_{\mathrm{LH}}(s') \geq 0$ and $\mathbf{a}_{\mathrm{LH}}(s') =0$. Also, for slice $s$, sum of the right-hand and left-hand parts of $\mathbf{n} - \mathbf{a}$ is equal to $Q^{\mathrm{max}}_s - 1$ and according to the definition (See \eqref{eq:979248}), we have $e_s =1$. As a result, we obtain $\mathbf{e}(\mathbf{n} - \mathbf{a}) = \mathbf{a}$ and therefore, $\Psi_{1} (\mathbf{n} - \mathbf{a}) = \Phi_{1} (\mathbf{n})$.

From \eqref{eq:237045} and \eqref{eq:300065}, we can write
\begin{equation} \nonumber
	\label{eq:999726}
		\Phi_{2} (\mathbf{n}) = \prod_{\substack{s' = 1, \\ \mathbf{n}_{\mathrm{RH}}(s') \neq 0, \\ e_{s'} \neq 1}}^{|\mathcal{S}|} \mathrm{M}_{s',u_{s'}}^{-1}(\mathbf{n}_{\mathrm{RH}}),
\end{equation}
and
\begin{equation} \nonumber
	\label{eq:999126}
	\begin{split}
	& \Psi_{2} (\mathbf{n} - \mathbf{a}) = \Big[\Phi_2(\mathbf{\mathbf{n} - \mathbf{a}})\Big]_{\mathbf{\mathbf{n} - \mathbf{a}}\rightarrow \mathbf{\mathbf{n} - \mathbf{a}} + \mathbf{e}(\mathbf{\mathbf{n} - \mathbf{a}})} = \Big[\Phi_2(\mathbf{\mathbf{n} - \mathbf{a}})\Big]_{\mathbf{\mathbf{n} - \mathbf{a}}\rightarrow \mathbf{\mathbf{n}}} \\ &=  \Big[\!\!\!\!\!\!\prod_{\substack{s' = 1, s' \neq s \\ \mathbf{n}_{\mathrm{RH}}(s') - \mathbf{a}_{\mathrm{RH}}(s') \neq 0, \\ e_{s'}\neq 1}}^{|\mathcal{S}|}\!\!\!\!\!\!\!\! \mathrm{M}_{s',u_{s'}}^{-1}(\mathbf{n}_{\mathrm{RH}} - \mathbf{a}_{\mathrm{RH}}) \Big]_{\mathbf{n} - \mathbf{a} \rightarrow \mathbf{n}} =  \prod_{\substack{s' = 1, s' \neq s \\ \mathbf{n}_{\mathrm{RH}}(s') \neq 0, \\ e_{s'} \neq 1}}^{|\mathcal{S}|} \mathrm{M}_{s',u_{s'}}^{-1}(\mathbf{n}_{\mathrm{RH}}),
	\end{split}
\end{equation}
where the latter equality is true because $\mathbf{n} - \mathbf{a}$ and $\mathbf{n}$ values are equal for all $s' \neq s$. Finally, by substituting the obtained values, we achieve
\begin{equation} \nonumber
	\label{eq:219126}
	\xi(\mathbf{n}, \mathbf{a}) = \mathrm{M}_{s,u_{s}}(\mathbf{n}_{\mathrm{RH}}).
\end{equation}
\clearpage
\section{Proof for Algorithm \ref{alg:beta}}
\label{app2}
Since we are using an iterative method, in iteration $k$, we can rewrite \eqref{eq:583071} for $s \in \mathcal{S}$ and $0 \leq \mathbf{n}_{\mathrm{RH}}(s) \textless Q^{\mathrm{max}}_s$ as follows
\begin{equation} \nonumber
	\hat{\beta}^{(k+1)}_{s, \mathbf{n}_{\mathrm{RH}}(s)} = \operatorname*{arg\,min}_{\hat{\beta}_{s, \mathbf{n}_{\mathrm{RH}}(s)}} \sum_{\mathrm{COND}(\mathbf{x},\mathbf{a})} \hat{\Phi}_{s, \mathbf{n}_{\mathrm{RH}}(s)}^{(k)}(\mathbf{x})(1 - \frac{\beta_{\mathbf{x},\mathbf{a}}}{\hat{\beta}_{s, \mathbf{n}_{\mathrm{RH}}(s)}})^2 .
\end{equation}
Now, we take the partial derivative with respect to $\hat{\beta}_{s, \mathbf{n}_{\mathrm{RH}}(s)}$. Considering that $\hat{\Phi}_{s, \mathbf{n}_{\mathrm{RH}}(s)}^{(k)}(\mathbf{x})$ has been obtained in previous iteration, it is independent of $\hat{\beta}_{s, \mathbf{n}_{\mathrm{RH}}(s)}$ and therefore
\begin{equation} \nonumber
	\begin{gathered}
		\frac{\partial}{\partial\hat{\beta}_{s, \mathbf{n}_{\mathrm{RH}}(s)}} \bigg( \sum_{\mathrm{COND}(\mathbf{x},\mathbf{a})} \hat{\Phi}_{s, \mathbf{n}_{\mathrm{RH}}(s)}^{(k)}(\mathbf{x})(1 - \frac{\beta_{\mathbf{x},\mathbf{a}}}{\hat{\beta}_{s, \mathbf{n}_{\mathrm{RH}}(s)}})^2\bigg)\\
		\begin{split}
			&=  \sum_{\mathrm{COND}(\mathbf{x},\mathbf{a})} \hat{\Phi}_{s, \mathbf{n}_{\mathrm{RH}}(s)}^{(k)}(\mathbf{x})\frac{\partial}{\partial\hat{\beta}_{s, \mathbf{n}_{\mathrm{RH}}(s)}}(1 - \frac{\beta_{\mathbf{x},\mathbf{a}}}{\hat{\beta}_{s, \mathbf{n}_{\mathrm{RH}}(s)}})^2\\
			&= \sum_{\mathrm{COND}(\mathbf{x},\mathbf{a})} \hat{\Phi}_{s, \mathbf{n}_{\mathrm{RH}}(s)}^{(k)}(\mathbf{x}) (\frac{2\beta_{\mathbf{x},\mathbf{a}}}{\hat{\beta}^{2}_{s, \mathbf{n}_{\mathrm{RH}}(s)}})(1 - \frac{\beta_{\mathbf{x},\mathbf{a}}}{\hat{\beta}_{s, \mathbf{n}_{\mathrm{RH}}(s)}}) =0. 
		\end{split}
	\end{gathered}
\end{equation}
Thereupon, we have 
\begin{equation} \nonumber
	\hat{\beta}^{(k+1)}_{s, \mathbf{n}_{\mathrm{RH}}(s)} = \frac{\sum_{\mathrm{COND}(\mathbf{x},\mathbf{a})} \hat{\Phi}_{s, \mathbf{n}_{\mathrm{RH}}(s)}^{(k)}(\mathbf{x})\beta_{\mathbf{x},\mathbf{a}}^2}{\sum_{\mathrm{COND}(\mathbf{x},\mathbf{a})} \hat{\Phi}_{s, \mathbf{n}_{\mathrm{RH}}(s)}^{(k)}(\mathbf{x})\beta_{\mathbf{x},\mathbf{a}}}.
\end{equation}